\documentclass[%
 reprint,
 superscriptaddress,
 amsmath,amssymb,
 aps,
prb,
floatfix,
longbibliography,
]{revtex4-1}

\usepackage{graphicx}
\usepackage{dcolumn}
\usepackage{bm}
\usepackage[normalem]{ulem}
\usepackage{hyperref}
\hypersetup{colorlinks = true, citecolor = blue, breaklinks = true}
\usepackage{placeins}

\usepackage{textcomp}
\usepackage{multirow}
\usepackage{xfrac}
\usepackage{units}

\newcommand{\angstrom}{\mbox{\normalfont\AA}}

\usepackage[version=4]{mhchem}

\begin{document}

\title{Combined theoretical and experimental study of the Moir\'e dislocation network at the \ce{SrTiO3}-\ce{(La,Sr)(Al,Ta)O3} interface}

\author{Chiara Ricca}
\affiliation{Department of Chemistry and Biochemistry, University of Bern, Freiestrasse 3, CH-3012 Bern, Switzerland}

\author{Elizabeth Skoropata}
\affiliation{Swiss Light Source, Paul Scherrer Institut, Forschungsstrasse 111, 5232 Villigen PSI, Switzerland}

\author{Marta D. Rossell}
\affiliation{Electron Microscopy Center, Empa, Swiss Federal Laboratories for Materials Science and Technology, {\"U}berlandstrasse 129, 8600 D\"ubendorf, Switzerland}

\author{Rolf Erni}
\affiliation{Electron Microscopy Center, Empa, Swiss Federal Laboratories for Materials Science and Technology, {\"U}berlandstrasse 129, 8600 D\"ubendorf, Switzerland}

\author{Urs Staub}
\affiliation{Swiss Light Source, Paul Scherrer Institut, Forschungsstrasse 111, 5232 Villigen PSI, Switzerland}

\author{Ulrich Aschauer}
\email{ulrich.aschauer@plus.ac.at}
\affiliation{Department of Chemistry and Biochemistry, University of Bern, Freiestrasse 3, CH-3012 Bern, Switzerland}
\affiliation{Department of Chemistry and Physics of Materials, University of Salzburg, Jakob-Haringer-Str. 2A, A-5020 Salzburg, Austria}

\date{\today}

\begin{abstract}
Recently a highly ordered Moir\'e dislocation lattice was identified at the interface between a \ce{SrTiO3} (STO) thin film and the \ce{(LaAlO3)_{0.3}(Sr2TaAlO6)_{0.7}} (LSAT) substrate. A fundamental understanding of the local ionic and electronic structure around the dislocation cores is crucial to further engineer the properties of these complex multifunctional heterostructures. Here we combine experimental characterization via analytical scanning transmission electron microscopy with results of molecular dynamics and density functional theory calculations to gain insights into the structure and defect chemistry of these dislocation arrays. Our results show that these dislocations lead to undercoordinated Ta/Al cations at the dislocation core, where oxygen vacancies can easily be formed, further facilitated by the presence of cation vacancies. The reduced \ce{Ti^{3+}} observed experimentally at the dislocations by electron energy-loss spectroscopy are a consequence of both the structure of the dislocation itself, as well as of the electron-doping due to oxygen vacancy formation. Finally, the experimentally observed Ti diffusion into LSAT around the dislocation core occurs only together with cation-vacancy formation in LSAT or Ta diffusion into STO.
\end{abstract}

\maketitle

\section{Introduction}

Complex transition metal perovskite oxides are a versatile class of materials with a wide spectrum of functional properties. They can be insulating, semiconducting, or metallic and show technologically relevant phenomena such as magnetism, ferroelectricity, or the more exotic high-temperature superconductivity and colossal magnetoresistance~\cite{Zubko2011, Lin2013, Bhattacharya2014, Arandiyan2021}. These properties are the result of a complex interplay of charge, orbital, spin, and lattice degrees of freedom~\cite{Zubko2011}, and depend strongly on strain and the defect chemistry~\cite{Kalinin2013, Chandrasena2017, Ricca2020}. The structural compatibility between different perovskites, allowing them to be stacked on top of each other, and the advances in deposition techniques enabled the fabrication of complex multifunctional heterostructures with relative ease. These heterostructures often give rise to interesting, novel, and unexpected physical phenomena emerging at the interface where materials with different structural and electronic properties meet: quasi-two-dimensional (2D) electron gas, colossal ionic conductivity, giant thermoelectric effect or resistance switching~\cite{Ohta2007, Barriocanal2008, Mannhart2010, Zubko2011, Zubko2012, Bhattacharya2014, Chen2017}.

Recently, a highly ordered Moir\'e lattice has been identified at the interface between \ce{SrTiO3} (STO) and \ce{(LaAlO3)_{0.3}(Sr2TaAlO6)_{0.7}} (LSAT) by high-resolution X-ray diffraction reciprocal space mapping~\cite{Burian2021}. A 30 nm thick film of STO was grown on LSAT (001) by pulsed laser deposition, followed by 12 h of annealing at 1200°C and ambient pressure. STO ($a_{\textrm{STO}}=3.905$\angstrom) and LSAT ($a_{\textrm{LSAT}}= 3.869$\angstrom) both have a cubic lattice with a small mismatch of 0.93\%. The high-temperature annealing allows the almost complete relaxation of the STO film, which results in the appearance of the Moir\'e pattern. This newly formed 2D network has a Moir\'e lattice constant of approximately 40~nm, corresponding to 106/107 unit cells of STO/LSAT, necessary to compensate for the small lattice mismatch between the two materials. Scanning transmission electron microscopy (STEM) images suggest that this periodicity is related to the appearance of a network of edge dislocations at the interface that has the same periodicity as the Moir\'e pattern.

The ability to form such ordered superlattices with a 2D network of line defects at complex perovskite oxide interfaces could be an emerging avenue to induce new interfacial functionality with unforeseen potential applications, such as 2D ferroics, 2D grid conductivity along the defect lines or ferroelectric three-dimensional (3D) vortex structures. These interfacial phenomena are the consequence of spin and charge interactions at the interface, which are in turn controlled by the local atomic arrangement. Hence, a detailed understanding of the atomic structure and electronic properties of the STO/LSAT  interface in presence of dislocations is key to interpreting the behavior of this heterostructure and to enhance its functional properties. With this goal, we report a combined experimental and theoretical study of the STO/LSAT heterostructure. STEM imaging along with chemical mapping by electron energy-loss (EELS) and energy-dispersive X-ray (EDX) spectroscopies have been combined with the results of molecular dynamics (MD) and density functional theory (DFT) calculations to deepen our understanding of the peculiarities of this system at the atomic scale and guide the design and optimization of these promising heterostructures.

\section{Methods}

\subsection{Computational details}

\subsubsection{Dislocation models}
%
\begin{figure}
	\centering
	\includegraphics[width=0.8\columnwidth]{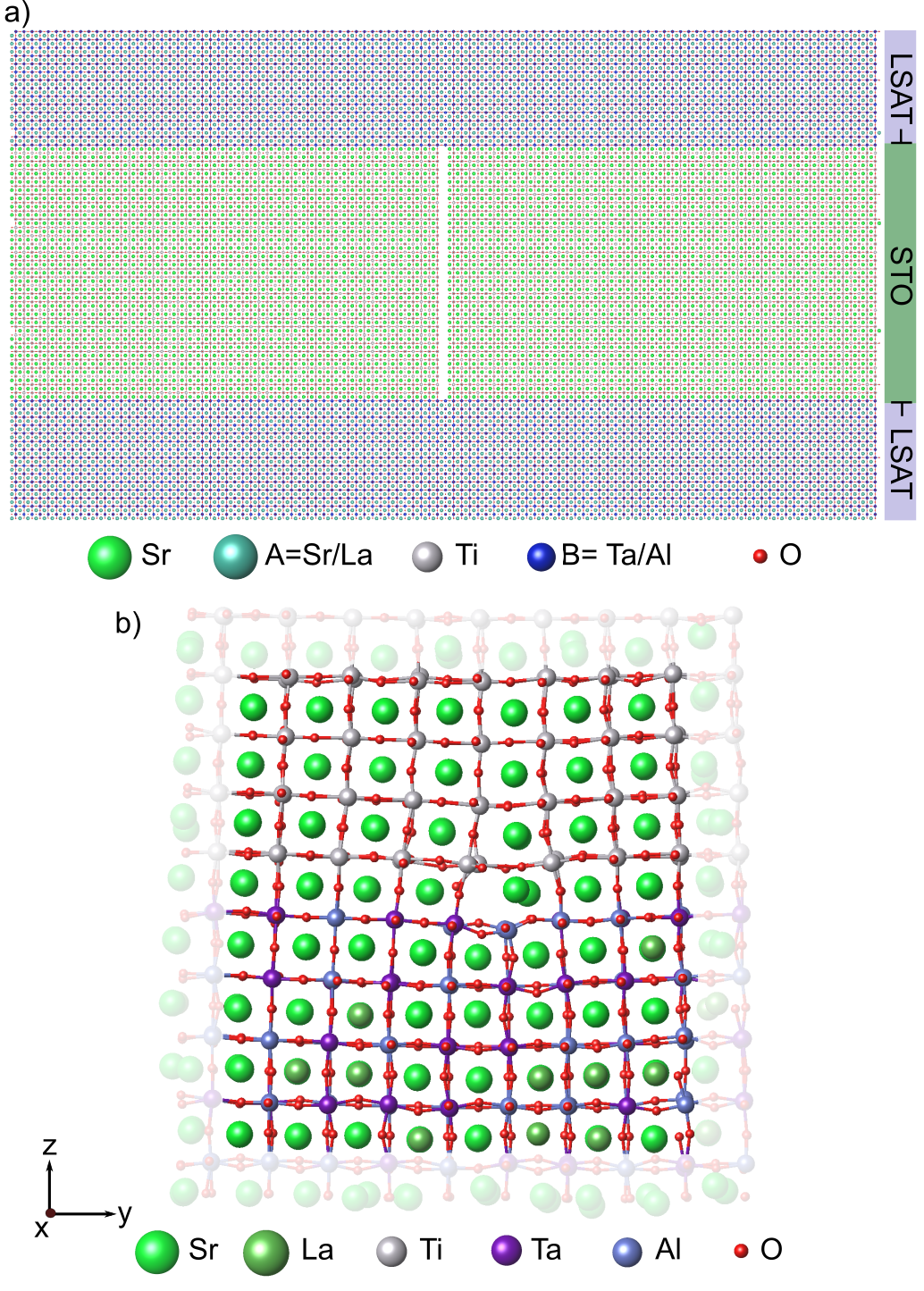}
	\caption{a) Initial structure of the ($2\times107\times60$) supercell used to simulate the 2D Moir\'e pattern at the interface of a STO thin film on a LSAT (001) substrate. b) Model of the dislocation at the STO/LSAT interface extracted from the MD simulations and relaxed using DFT. The shaded atoms are fixed during DFT geometry optimization to impose elastic boundary conditions.}
	\label{fig:structures}
\end{figure}
The simulation of these dislocation arrays is complicated by their non-periodic nature along the interface normal and by their large ordering period along the interface. An appropriate description along both of these directions is however crucial to accurately determine the long-range strain field associated with the dislocation. In order to obtain a reliable description of the 2D Moir\'e pattern at the interface of a thin film of STO on LSAT (001), we employed the simulation setup schematically shown in Fig.~\ref{fig:structures}, which is a combination of the simulation approaches suggested in Refs.~\cite{HIREL2012329, sun2015, Marrocchelli2015}. We started by creating two dislocations with opposite Burgers vectors in a simulation box with dimensions $\sim\unit[0.75]{nm}\times\unit[41.50]{nm}\times\unit[23.51]{nm}$  (approximately 64'000 atoms) corresponding to a $2\times107\times60$ supercell of a 5-atom cubic perovskite unit cell. The number of cells along the $y$ axis allows us to reproduce the observed periodicity of the Moir\'e superlattice, while along the $z$ axis we stacked 30 STO on 30 LSAT layers. This means that the two dislocation cores are separated by about \unit[11.76]{nm} along the $z$ axis which is sufficient to significantly suppress interactions between them. A missing plane is then created in the middle of the cell by removing one \ce{SrO} and one \ce{TiO2} plane (see Fig.~\ref{fig:structures}a). The system is relaxed via classical molecular dynamics (see below for details), during which the missing plane heals, forming two dislocations with opposite Burgers vectors at the upper and lower interface. 

DFT calculations (see below for details) are then performed on a cluster model created from the final MD structure by extracting a section of $2\times10\times10$ STO/LSAT unit cells around the bottom dislocation core (see Fig.~\ref{fig:structures}b). This cell contains 950 atoms and 5 STO and 5 LSAT layers along the $z$ axis. For the LSAT substrate, 18 La and 82 Sr, as well as 59 Al and 41 Ta atoms, were randomly introduced at the A and B sites, respectively, to obtain a charge-neutral LSAT layer with a composition (\ce{La_{0.2}Sr_{0.8}Al_{0.4}Ta_{0.6}O6}) similar to experiment (\ce{La_{0.3}Sr_{0.7}Al_{0.3}Ta_{0.7}O6}). The cluster was periodically repeated along the dislocation line only ($x$ axis), while in the perpendicular directions only the periodicity of the electrostatic potential was imposed. Furthermore, atoms in the boundary region were kept fixed throughout the simulation to impose the elastic boundary conditions of the interface and dislocation environment. Within this approach, it is possible to avoid artifacts due to the interaction of the dislocation with its images in neighboring cells, thus isolating the dislocation core. The dislocation core is embedded in the correct long-range elastic field, allowing, at the same time, to minimize any spurious effect due to long-range electrostatic fields~\cite{HIREL2012329}.

Neutral vacancies (\ce{V_X}) were created by removing one X (X = O, La, Al, or Ta) atom from the DFT dislocation model. Ti substitution at Al (\ce{Ti_{Al}}) or Ta (\ce{Ti_{Ta}}) sites and Ta substitution of Ti atoms (\ce{Ta_{Ti}}) were also taken into account. Different possible vacancy and substitutional defect configurations, involving different sites around the dislocation, were considered.

\subsubsection{Molecular dynamics}

The LAMMPS~\cite{PLIMPTON19951} code was used to perform the classical MD simulations. Interatomic interactions were described using a non-polarizable rigid-ion model consisting of long-range electrostatic interactions between the nuclei and a short-range Buckingham potential with a cut-off radius of 12~\angstrom:
\begin{equation}
	V_{ij} = \frac{q_i q_j}{4 \pi \epsilon _0 r_{ij}} + A_{ij}  e^{- \frac{r_{ij}}{\rho _{ij}}} - \frac{C_{ij} }{r_{ij}^6}\,,
	\label{eq:inteatomicpotential}
\end{equation}
where $q_{i/j}$ are the atomic charges, $r_{ij}$ their separation and $A_{ij}$, $\rho_{ij}$ and $C_{ij}$ the parameters of the short-range potential as reported in Table~\ref{tbl:intpot}. The short-range parameters were derived by starting from the Lewis and Catlow~\cite{Lewis_1985} set and using the GULP code~\cite{gulp1, gulp2, gulp3} to fit lattice parameters and elastic constants of STO and LSAT derived from DFT calculations (see below). For LSAT, we chose to describe the interaction between O and A/B dummy atoms representing the average properties of the La/Sr and Al/Ta atoms occupying the A and B sites respectively. MD simulations were performed in the canonical (NVT) ensemble with a Nose-Hoover thermostat and barostat. The system was allowed to relax first at 50 K for 50 ps, then at 1473 K for 130 ps before the temperature was reduced again to 50 K over 120 ps. In order to drain the kinetic energy released during closing the missing plane in a controlled way, a viscous damping force with a coefficient $\gamma=\unit[1.0]{eV\cdot ps/\angstrom^2}$ is applied to all atoms.
\begin{table}
	\caption{Parameters of the Coulomb-Buckingham potential for STO and LSAT.}
	\begin{tabular*}{\columnwidth}{@{\extracolsep{\fill}}llrrrr}
		\hline
		\hline
    $i$ & $j$ & $q_i$ (e) & A$_{ij}$  (eV) & $\rho_{ij}$ (\angstrom) & C$_{ij}$ (eV/\angstrom$^{-6}$) \\
		 \hline
		Sr & O & 2.00 & 1324.77 & 0.3008 & 0.00 \\
		Ti & O & 4.00 & 762.26 & 0.4014 & 0.00 \\
		A & O & 2.18 & 2018.14 & 0.2876 & 0.00 \\
		B & O & 3.82 & 867.03 & 0.3828 & 0.00 \\
		O & O & -2.00 & 22764.30 & 0.1490 & 31.15 \\ 
		\hline
	\end{tabular*}
	\label{tbl:intpot}
\end {table}

DFT calculations to determine lattice parameters and elastic constants used in potential fitting were performed with the VASP code~\cite{Kresse:1993ty, Kresse:1994us, Kresse:1996vk, Kresse:1996vf}. We used the PBE~\cite{PBE} exchange-correction functional together with PAW~\cite{Blochl:1994uk, Kresse:1999wc} potentials with La(5\textit{s}, 5\textit{d}, 5\textit{p}, 6\textit{s}), Sr(4\textit{s}, 4\textit{p}, 5\textit{s}), Al(3\textit{s}, 3\textit{p}), Ta(5\textit{p}, 5\textit{d}, 6\textit{s}) and O(2\textit{s}, 2\textit{p}) valence electrons and a plane-wave cutoff of 550 eV. For LSAT we used a $\unit[7.76]{\angstrom}\times\unit[10.97]{\angstrom}\times\unit[10.97]{\angstrom}$ $2\times2\sqrt{2}\times2\sqrt{2}$ supercell of the 5-atom cubic unit cell with composition \ce{La3Sr13Al9Ta7O48}, reciprocal space of which was sampled with a $4\times4\times4$ Monkhorst-Pack~\cite{monkhorst1976special} mesh, while a $8\times8\times8$ mesh was used for the 5-atom STO unit-cell ($\unit[3.90]{\angstrom}\times\unit[3.90]{\angstrom}\times\unit[3.90]{\angstrom}$). Structures were relaxed until forces converged below $\unit[10^{-4}]{eV/\angstrom}$ before elastic constants were determined using a central finite-differences approach with a step-size of \unit[0.015]{\angstrom}.

\subsubsection{DFT calculations}

The DFT calculations on the cluster model were performed with the CP2K program package~\cite{Kuehne2020} using the PBE~\cite{PBE} exchange-correlation functional. The norm-conserving Goedecker-Teter-Hutter (GTH) pseudopotentials~\cite{Goedecker1996} together with the GTH double-$\zeta$ polarized molecularly optimized basis sets~\cite{VandeVondele2007} and an energy cutoff of \unit[750]{Ry} were applied. The convergence criterion for the self-consistent field method was set to $\unit[10^{-6}]{Ha}$, while atomic positions were relaxed within a force threshold of $\unit[10^{-3}]{eV/\angstrom}$.

The defect formation energy ($E_\textrm{f}$) of a neutral defect was calculated as described in Ref.~\cite{freysoldt2014first}:
\begin{equation}
	E_\textrm{f}(\mu_i) = E_\textrm{tot,def} - E_\textrm{tot,stoic} - \sum_i n_i \mu_i \,
	\label{eq:formenerg}
\end{equation}
where $E_\textrm{tot,def}$ and $E_\textrm{tot,stoic}$ are the DFT total energies of the defective system and of the stoichiometric cell, respectively. $n_i$ indicates the number of atoms of a certain specie $i$ that is added ($n_i > 0$) or removed ($n_i < 0$) from the supercell to form the defect, while $\mu_i$ is the species' chemical potential. For simplicity, we used the atomic energy of the corresponding reference phase for each element (metal La, Ta, Ti, Al and molecular \ce{O2}) as the chemical potential. 

\subsection{Experimental methods}

\subsubsection{Scanning transmission electron microscopy}
Electron transparent samples for STEM investigations were produced in cross-section geometry using a FEI Helios 660 G3 UC dual-beam focused (Ga) ion beam instrument operated at 30 and 5 kV, after deposition of C and Pt protective layers. High-angle annular dark-field (HAADF) STEM, electron energy-loss spectroscopy (EELS), and energy-dispersive X-ray spectroscopy (EDX) were carried out using a probe aberration-corrected FEI Titan Themis microscope operated at 300 kV and equipped with a SuperX EDX system and a CEFID energy filter in combination with an ELA direct electron detector (for details see, Ref.~\cite{Ruiz22}). For the HAADF-STEM data acquisition, a probe convergence semi-angle of 26 mrad was set and the inner angle of the annular semi-detection range was 171 mrad. The EELS data were obtained with a collection semi-angle of 35 mrad yielding an effective collection angle of about 29.5 mrad considering the energy range of the spectra.

A quantitative analysis of the lattice distortions at the dislocation cores was performed by means of peak-pair analysis (PPA), fitting the peaks corresponding to the atomic columns in the high-resolution HAADF-STEM images as described in Ref.~\cite{Campanini2018}. In particular, we analyzed the structural distortions by measuring the distance between the peaks corresponding to the atomic columns of the A sublattice.

\section{Results and discussion\label{sec:results}}

\subsection{Structure and strain field at the dislocation core}
%
\begin{figure}
	\centering
	\includegraphics[width=0.8\columnwidth]{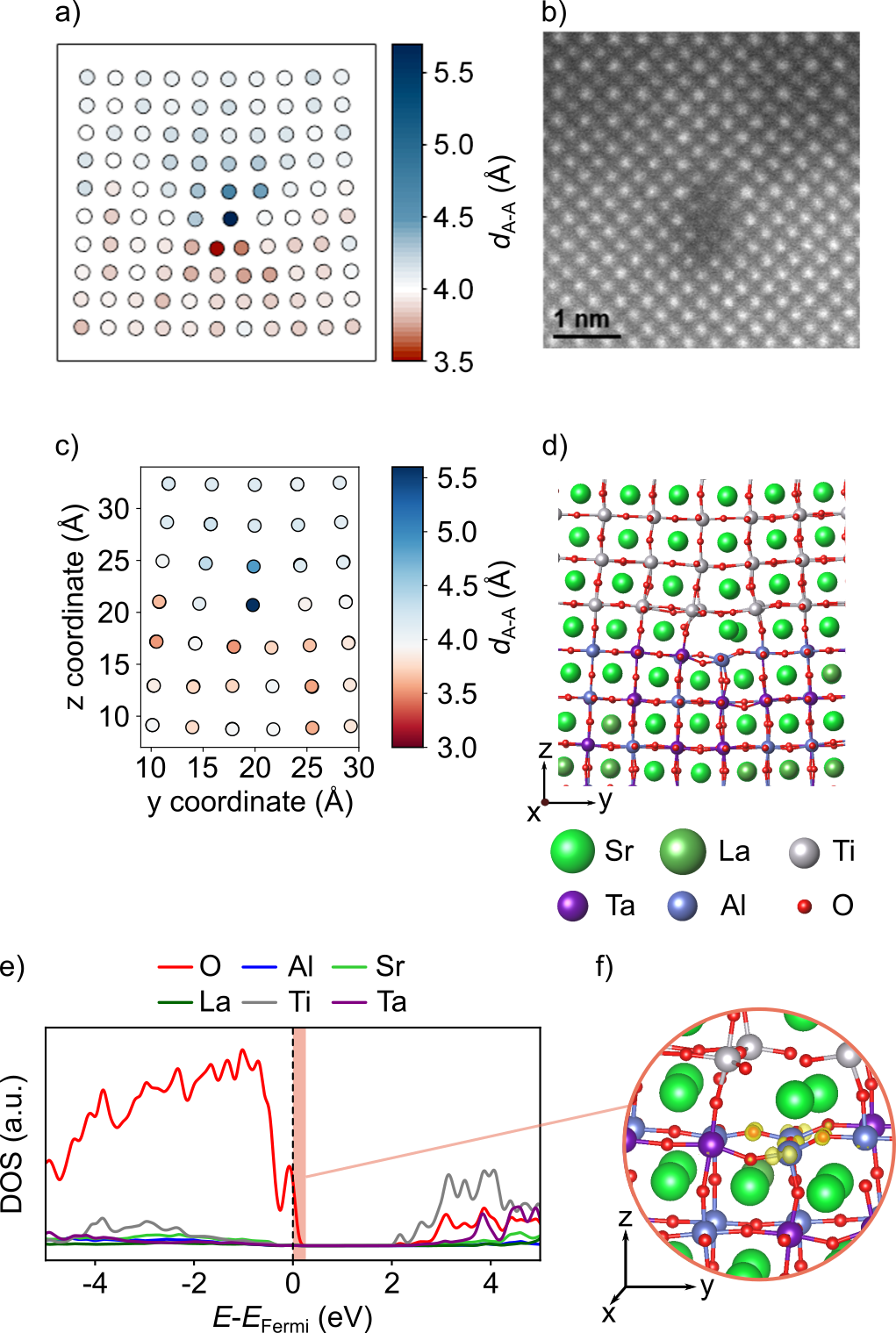}
	\caption{a) Map of the interatomic distances (in \angstrom) for the A sublattice along the horizontal direction as obtained by PPA from the HAADF-STEM image in panel b). c) Map of horizontal interatomic A-A distances extracted from the DFT relaxed dislocation core structure shown in panel d). Circles are located at the midpoint of A-A pairs and color-coded according to the distance between that pair. e) Electronic density of states (DOS) projected on the atoms at the dislocation core for the stoichiometric dislocation model. The vertical dashed line indicates the position of the Fermi level. f) Charge density isosurface ($\unit[10^{-2}]{e/\angstrom^3}$) in the energy range highlighted in red in e).}
	\label{fig:strainfield_dos}
\end{figure}
%
%
%
As already discussed in Ref.~\cite{Burian2021}, high-resolution STEM images of the annealed STO/LSAT samples reveal the presence of a highly ordered arrangement of edge dislocations with periodicity equivalent to the one of the Moir\'e lattice observed by high-resolution X-ray diffraction reciprocal space mapping. These dislocations form where the local structural mismatch between the two materials is largest (see Fig.~\ref{fig:strainfield_dos}b). In the present work, the HAADF-STEM images have been used to map the strain field around the dislocation core as reported in Fig.~\ref{fig:strainfield_dos}a). The strain field is, indeed, highly localized around the dislocation core, where two different strain regions can be identified: a tensile strain region in the STO film and a compressive strain region in the LSAT layer.

Figure~\ref{fig:structures}d) shows the final structure of the dislocation core obtained after DFT geometry optimization starting from the structure extracted from the MD simulations. The strain field associated with this model was computed by extracting A-A distances parallel to the interface as shown in Figure~\ref{fig:structures}c. In this case, the largest strains are also observed at the dislocation core with an average tensile/compressive strain in the STO/LSAT layer, respectively. Overall, the strain-field maps derived from experiments and DFT calculations show a qualitative agreement, allowing us to confidently use the DFT model to understand the structural properties of the dislocation core at the atomic level.

As shown in Figure~\ref{fig:strainfield_dos}d, highly strained Ti-O-Ti bonds are established across the missing planes, leaving the B cations of the extra \ce{Ta/AlO2} plane undercoordinated at the dislocation core. The missing Al-O bonds at the STO/LSAT interface induce hole doping of the system, as can be seen from the density of states (DOS) projected on the atoms at the dislocation core (Fig.~\ref{fig:strainfield_dos}e), where the Fermi level crosses the top of the valence band, which is formed by the O-2$p$ states of the O atoms bonded to the undercoordinated B cations at the dislocation core (Fig.~\ref{fig:strainfield_dos}f).

Despite the overall qualitative agreement, the dislocation core imaged by STEM shows contrast differences larger than expected from the structural and electronic changes in the DFT model, which could result from the DFT model being constructed from stoichiometric STO and LSAT, while point defects, such as oxygen and cation vacancies or substitutional defects, may be present in experiment.

\subsection{Defect chemistry of the dislocation core}

\subsubsection{Oxygen vacancies}
%
\begin{figure}
	\centering
	\includegraphics[width=0.8\columnwidth]{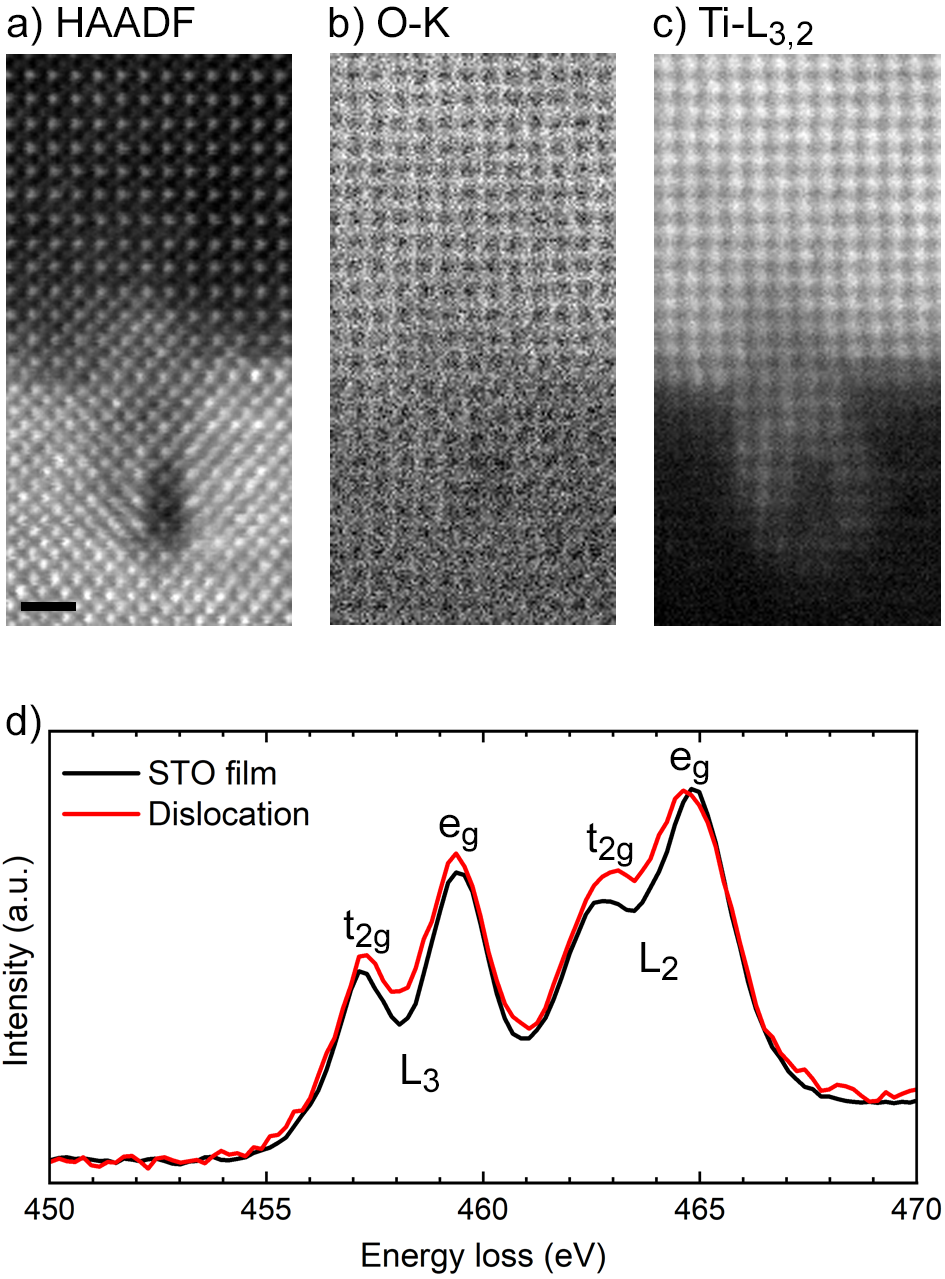}
	\caption{a) HAADF-STEM image of a dislocation at the STO/LSAT interface and corresponding atomic-resolution maps of the b) O-K and c) Ti-L$_{3,2}$ excitation edges. The scale bar is 1 nm. d) Representative Ti-L$_{3,2}$ EELS spectra of the STO film and the surroundings of the dislocation core. The spectra have been normalized to the L$_{2}$ e$_{g}$  peak height for clarity.}
	\label{fig:eels}
\end{figure}
EELS applied to the STO/LSAT interface can be used to obtain information on both the O and Ti states from the O-K (O $1s$ $\rightarrow$ $2p$) and the Ti-L$_{3,2}$ (Ti $2p$ $\rightarrow$ $3d$) core edges and could thus provide information about the presence of oxygen vacancies at the interface between the two oxides. The O-K edge map of Fig.~\ref{fig:eels}b) appears to be darker at the dislocation core, pointing to the presence of oxygen vacancies in this region. At the O-K edge, unoccupied O $p$-density of states in the presence of a core hole are probed and, thus, its intensity is expected to be reduced in the presence of oxygen vacancies that are, generally, electron donors in transition metal oxides~\cite{Muller2004}.

\begin{figure}
	\centering
	\includegraphics[width=0.8\columnwidth]{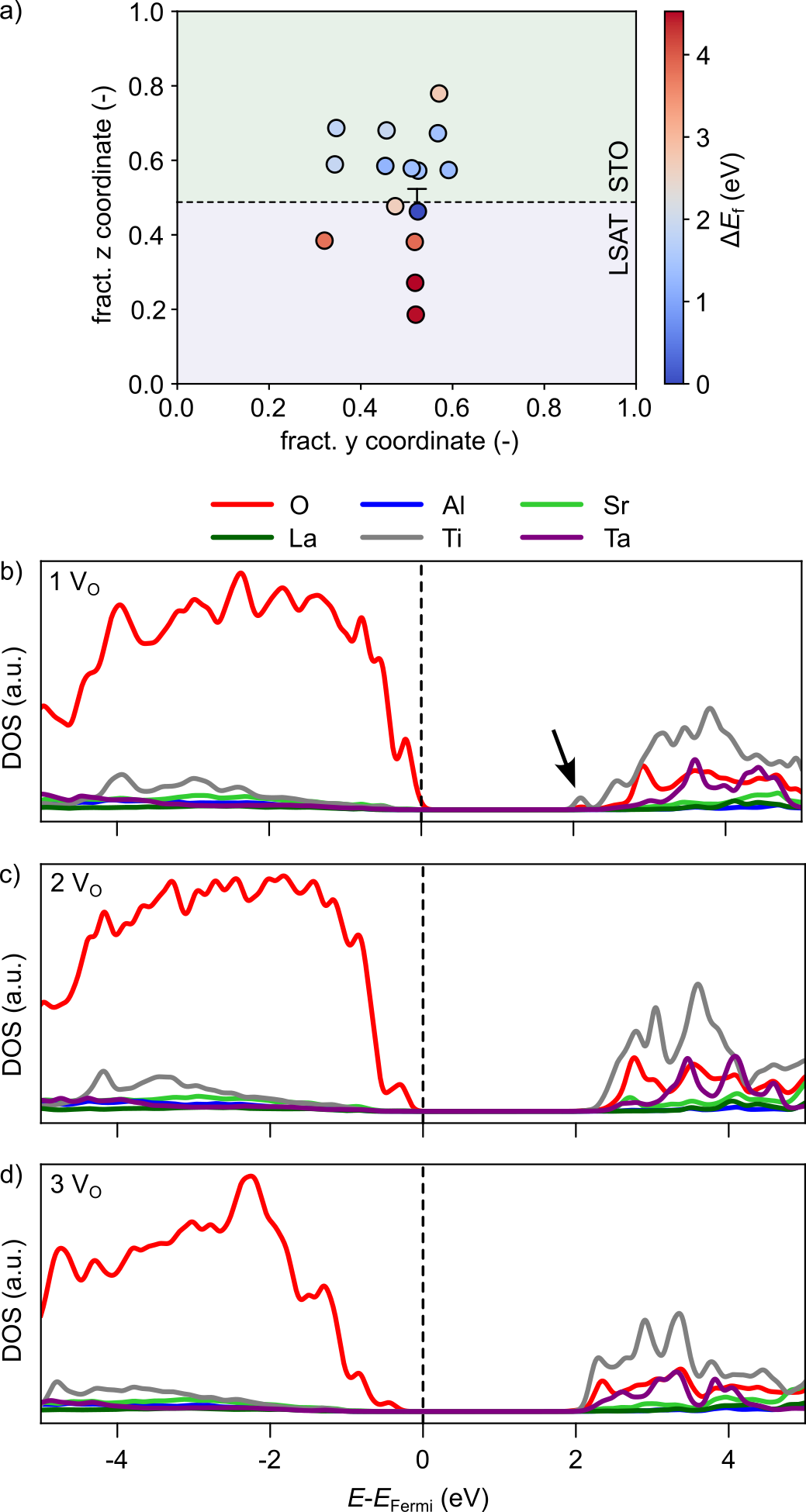}
	\caption{a) Color map of the relative \ce{V_O} formation energy ($\Delta E_\textrm{f}$) around the dislocation core. Electronic density of states (DOS) projected on the atoms at the dislocation core for the STO/LSAT interface model containing b) one, c) two, and d) three V$_{\textrm{O}}$ in the most stable configurations. The vertical dashed line indicates the position of the Fermi level.}
	\label{fig:VO_A_pdos}
\end{figure}
To investigate the formation of oxygen vacancies (\ce{V_O}) around the dislocation using DFT, we considered for simplicity a single neutral \ce{V_O} created at different oxygen sites in our STO/LSAT dislocation model. The map reported in Fig.~\ref{fig:VO_A_pdos}a is color-coded according to the \ce{V_O} formation energy of each \ce{V_O} site relative to the most stable site ($\Delta E_\textrm{f}$). The  energetically most favorable site to form a \ce{V_O} is at the dislocation core for the O atoms of the extra \ce{Ta/AlO2} plane closest to the STO/LSAT interface. This can be ascribed to the removal of an undercoordinated O atom to form this defect. It was already observed, that similarly to what happens at surfaces and grain-boundaries, also at the dislocation core, undercoordination can result in a lowering of $E_\textrm{f}$~\cite{Marrocchelli2015}. Furthermore, as seen above, the DOS of the stoichiometric dislocation model is characterized by hole states right above the Fermi level and localized on the undercoordinated O atoms at the dislocation core (Fig.~\ref{fig:strainfield_dos}e). These hole states can be filled by the two extra electrons left in the lattice upon formation of a neutral \ce{V_O} (Fig.~\ref{fig:VO_A_pdos}b). The formation energy gradually increases with increasing distance from the dislocation core, with larger values in LSAT compared to STO, which is easier to reduce, as shown by the computed \ce{V_O} formation energies in the two bulk materials (STO: 3.90~eV, LSAT: 5.04~eV). It also agrees with tensile strain, as in the STO layer, favoring neutral \ce{V_O} formation based on chemical expansion arguments~\cite{Aschauer2015}. These results suggest that \ce{V_O} have the tendency to segregate to the dislocation core and in particular to the STO side of the interface, in agreement with information obtained from the O-K edge EELS spectral map.

The Ti-L$_{3,2}$ map of Fig.~\ref{fig:eels}c shows that Ti atoms tend to diffuse into LSAT. Interestingly, the Ti-L$_{3,2}$ EELS spectra acquired in the STO film far away from the interface and around the dislocation core show striking differences (Fig.~\ref{fig:eels}d). Both spectra are composed of two main features, namely the L$_{3}$ and L$_{2}$ edges, separated by about $\sim$5.5 eV due to the spin-orbit splitting of the Ti $2p$ core hole into $2p_{3/2}$ and $2p_{1/2}$ states. Besides, these edges are further subdivided into two peaks, the t$_{2g}$ and e$_{g}$ peaks, by the strong octahedral crystal-field splitting arising from the surrounding oxygen atoms. In particular, the spectrum acquired at the dislocation is characterized by a relative increase in the spectral weight of the Ti-L$_{3,2}$ t$_{2g}$ peaks, especially in the higher-energy L$_{2}$ edge, compared to the spectrum of the STO film. Previously, such spectral changes were related to the presence of \ce{Ti^{3+}} in nominally \ce{Ti^{4+}}-based perovskite oxides \cite{Abbate1991}. Aside, the energy shift to lower energies of the Ti-L$_{3,2}$ edge observed in the dislocation spectrum with respect to the spectra obtained in the STO film suggests the presence of reduced \ce{Ti^{3+}} species in the dislocation core compared to \ce{Ti^{4+}} in the STO film~\cite{Muller2004}.

This change in the oxidation state is consistent with electron doping due to neutral \ce{V_O} at the dislocation core. Unfortunately, our DFT results did not show a complete reduction of Ti atoms at or around the dislocation core. Indeed, the DOS in Fig.~\ref{fig:VO_A_pdos}b-d do not show any filled localized state with Ti-$3d$ character, inherent with reduction of Ti atoms to \ce{Ti^{3+}} even in presence of three \ce{V_O}, positioned, for simplicity, at the most stable sites according to Fig.~\ref{fig:VO_A_pdos}a. The absence of DOS features related to Ti reduction could be due to the limitations of our DFT method. It is well known that standard semi-local DFT functionals, such as PBE, fail in localizing charge on Ti atoms neighboring a \ce{V_O} for small bulk STO cells~\cite{Ricca2020_sto}. Furthermore, even more accurate hybrid functionals do not necessarily lead to Ti reduction when \ce{V_O} are formed in large STO supercells, such as the one used in the present work~\cite{Ricca2020_sto}.

\begin{figure}
	\centering
	\includegraphics[width=0.8\columnwidth]{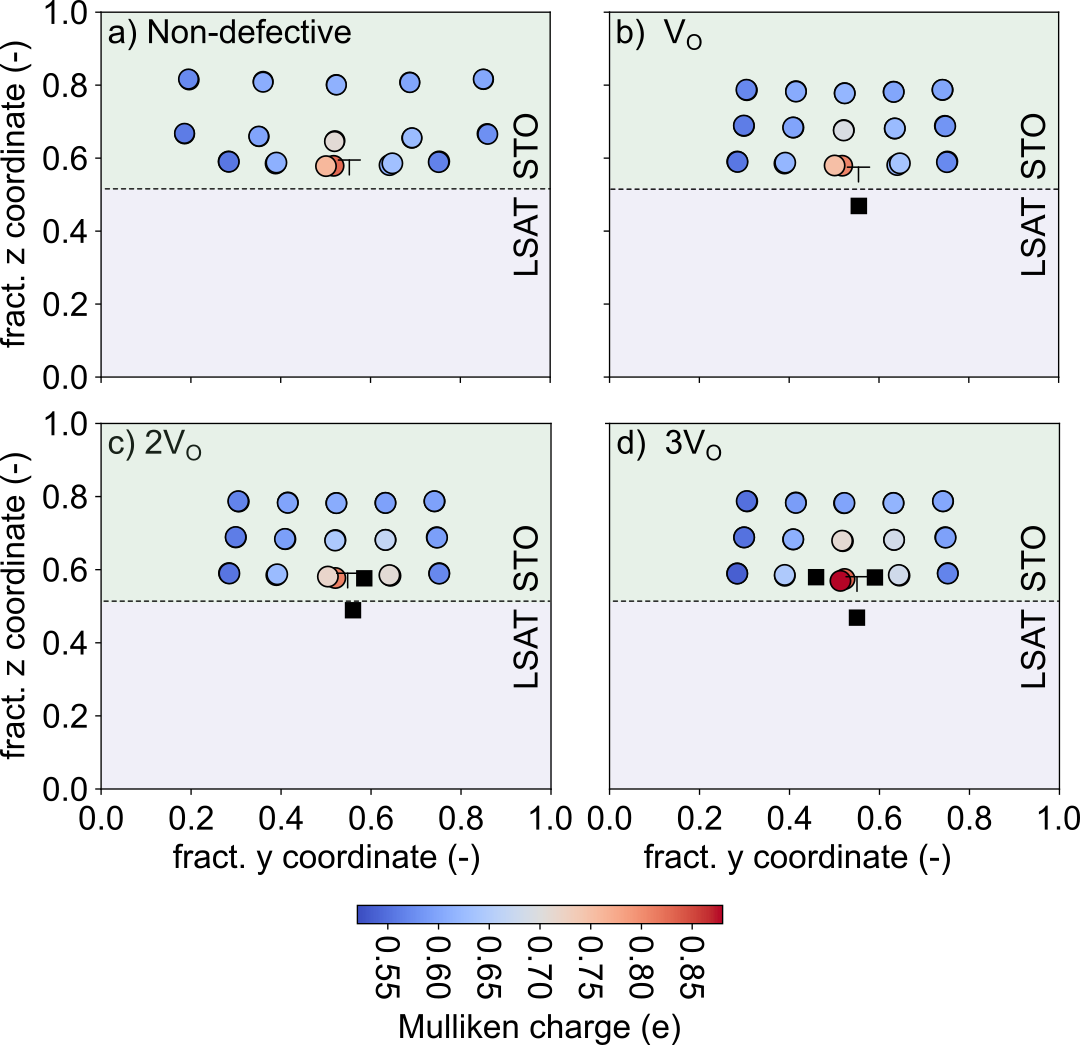}
	\caption{Mulliken charges for the Ti atoms in a) the stoichiometric dislocation model and in presence of b) one, c) two, and d) three \ce{V_O}. The black squares indicate the position of the \ce{V_O}.}
	\label{fig:mulliken}
\end{figure}
However, with increasing the number of \ce{V_O}, we observe a small peak with Ti-$3d$ character at the bottom of the CB (see arrow in Fig.~\ref{fig:VO_A_pdos}b) that is present for a single \ce{V_O} to disappear, while the states with O-$2p$ character at the top of the valence band are lowered in energy. This suggests a filling of Ti states for an increasing number of \ce{V_O} and hence increasing electron doping. This is supported by an analysis of the Mulliken charges for Ti atoms. Already in the stoichiometric dislocation model, Ti ions close to the dislocation core are more reduced compared to the rest of the Ti sites in STO (Fig.~\ref{fig:mulliken}a). Creation of the first \ce{V_O} at the dislocation core, when the dislocation hole state is only partially filled, does not significantly alter the charge of these Ti atoms. These charges, however, increase for 2 or 3 \ce{V_O} in STO around the dislocation (Figs.~\ref{fig:mulliken}b-d)). These results suggest that the \ce{Ti^{3+}} detected by EELS in the vicinity of the dislocation stems from both the presence of the dislocation and the formation of \ce{V_O}.

\subsubsection{Cation vacancies}
%
\begin{figure}
	\centering
	\includegraphics[width=0.8\columnwidth]{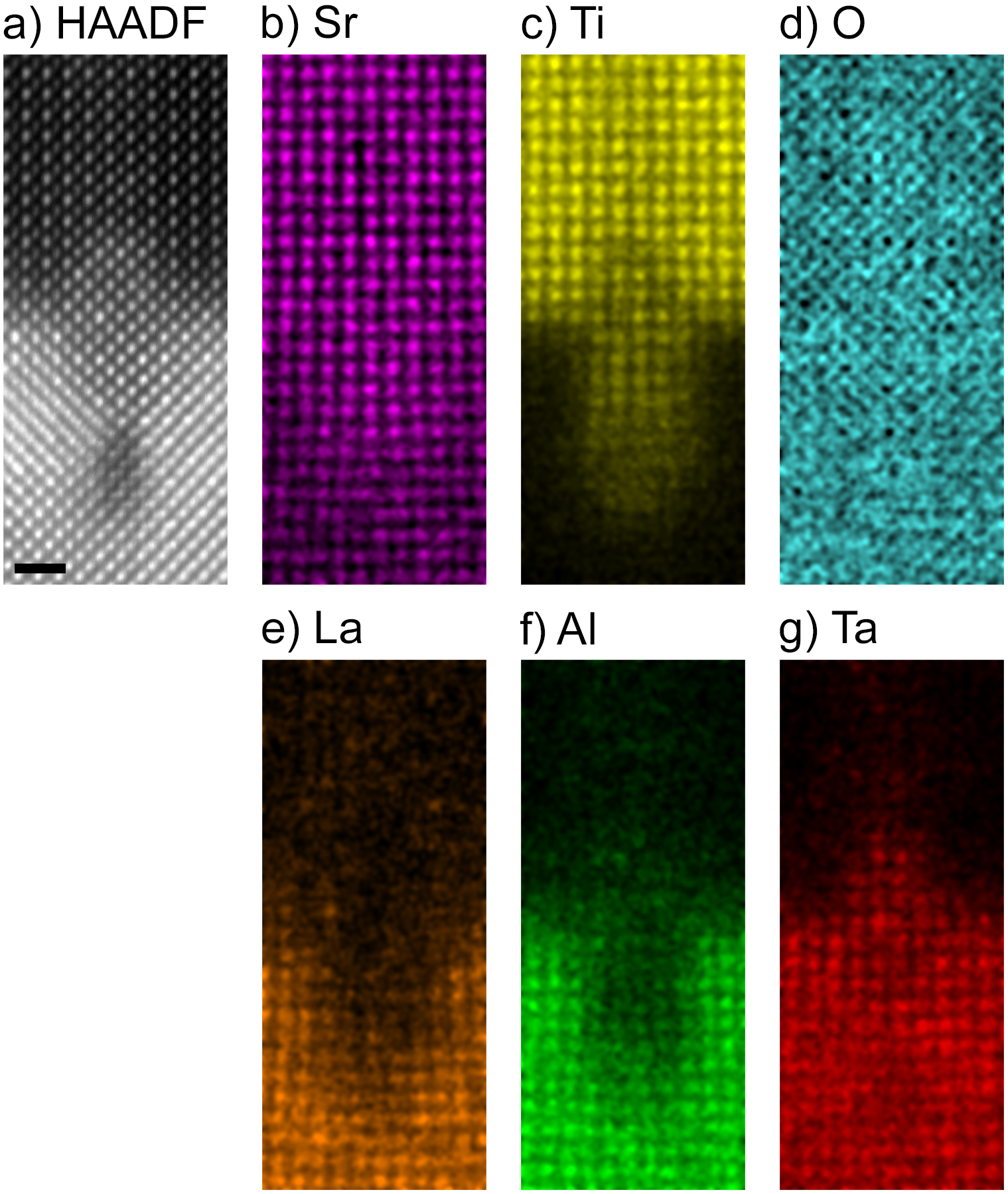}
	\caption{a) HAADF-STEM image of a misfit dislocation at the STO/LSAT interface and b-g) corresponding elemental maps of Sr, Ti, O, La, Al and Ta calculated from an EDX spectrum image using the Sr-K, Ti-K, O-\ce{K\alpha_{1}}, La-L, Al-\ce{K\alpha_{1}} and Ta-L lines, respectively. The scale bar is 1 nm.}
	\label{fig:edx}
\end{figure}

To further characterize the defect chemistry of the STO/LSAT interface, we performed chemical mapping with atomic resolution using EDX. The positions of the atomic columns visualized in the EDX maps, shown in Fig.~\ref{fig:edx}b-g) for Sr, Ti, O, La, Al, and Ta, reveal the chemical structure at the STO/LSAT interface at and around the dislocation core: cation vacancies (especially \ce{V_{Al}}, \ce{V_{La}}, and \ce{V_{Ta}}) are formed around the dislocation core, which are (partially) filled by Ti diffusing from STO into LSAT. At the same time, a partial substitution of Ti by Ta takes place in the STO film above the dislocation.

EDX mapping suggests that a combination of different defects accompanies dislocation formation that is driven by strain relaxation. It is, however, impractical to compute these defects simultaneously due to the size of the model and the large number of defect types and configurations that would have to be taken into account. We thus try to identify guiding rules for point defect formation in the vicinity of the dislocation core at the STO/LSAT interface by considering the most relevant defect types separately.

\begin{figure}
	\centering
	\includegraphics[width=0.8\columnwidth]{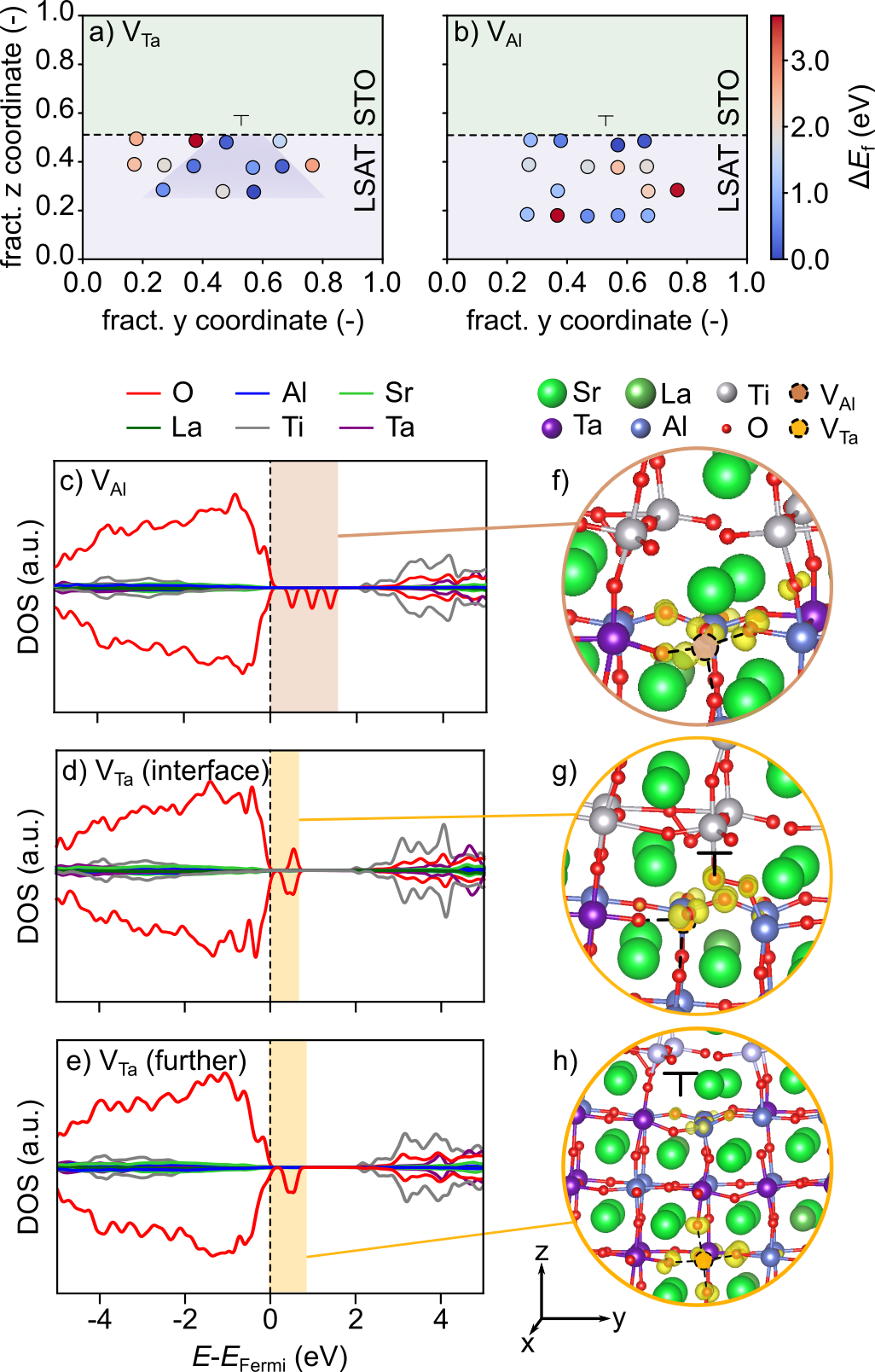}
	\caption{Color map of the relative a) \ce{V_{Ta}} and b) \ce{V_{Al}} formation energy ($\Delta E_\textrm{f}$) around the dislocation core. The energy of the most stable defect is used as a reference in each plot. The shaded trapezoidal area in a) indicates where \ce{V_{Ta}} tend to form/segregate to. Electronic density of states (DOS) projected on the atoms at the dislocation core containing c) one \ce{V_{Al}} and d) one \ce{V_{Ta}}, each at the most stable site at the STO/LSAT interface or e) away from the interface in LSAT. Panels f-h) show charge density isosurfaces ($\unit[10^{-2}]{e/\angstrom^3}$) in the energy range of the defect states highlighted in brown or orange in the DOS.}
	\label{fig:VB_pdos}
\end{figure}

We start by investigating the formation of a single neutral Ta (\ce{V_{Ta}}) and Al (\ce{V_{Al}}) vacancy, for which experiment shows Ta atoms to diffuse into STO, while Al diffuses out of dislocation cores that appear elementally hollow. The map of relative formation energies in Fig.~\ref{fig:VB_pdos}a indicates that one of the most stable \ce{V_{Ta}} configurations is at the STO/LSAT interface, closest to the dislocation core. Interestingly, all other \ce{V_{Ta}} have much larger formation energies (by 1.2~eV) and are thus much less favorable to form. More generally, we observe that \ce{V_{Ta}} positions with low $E_\textrm{f}$ are located in a trapezoidal area below the dislocation core (see Fig.~\ref{fig:VB_pdos}a). This peculiar profile matches with the magnitude of the compressive strain in the LSAT region (Fig.~\ref{fig:strainfield_dos}c), which is known to lower the formation of cation vacancies~\cite{Aschauer2015}. This result suggests that \ce{V_{Ta}} strongly favor a trapezoidal area below the dislocation core, in agreement with the EDX maps and the STEM image contrast (Fig.~\ref{fig:strainfield_dos}b). Instead, \ce{V_{Al}} formation is less sensitive to strain, consistent with the smaller radius of the \ce{Al^{3+}}, favorable sites being both at the STO/LSAT interface close to the dislocation core, as well as further into the LSAT substrate, in agreement with EDX mapping. 

\begin{figure}
	\centering
	\includegraphics[width=0.8\columnwidth]{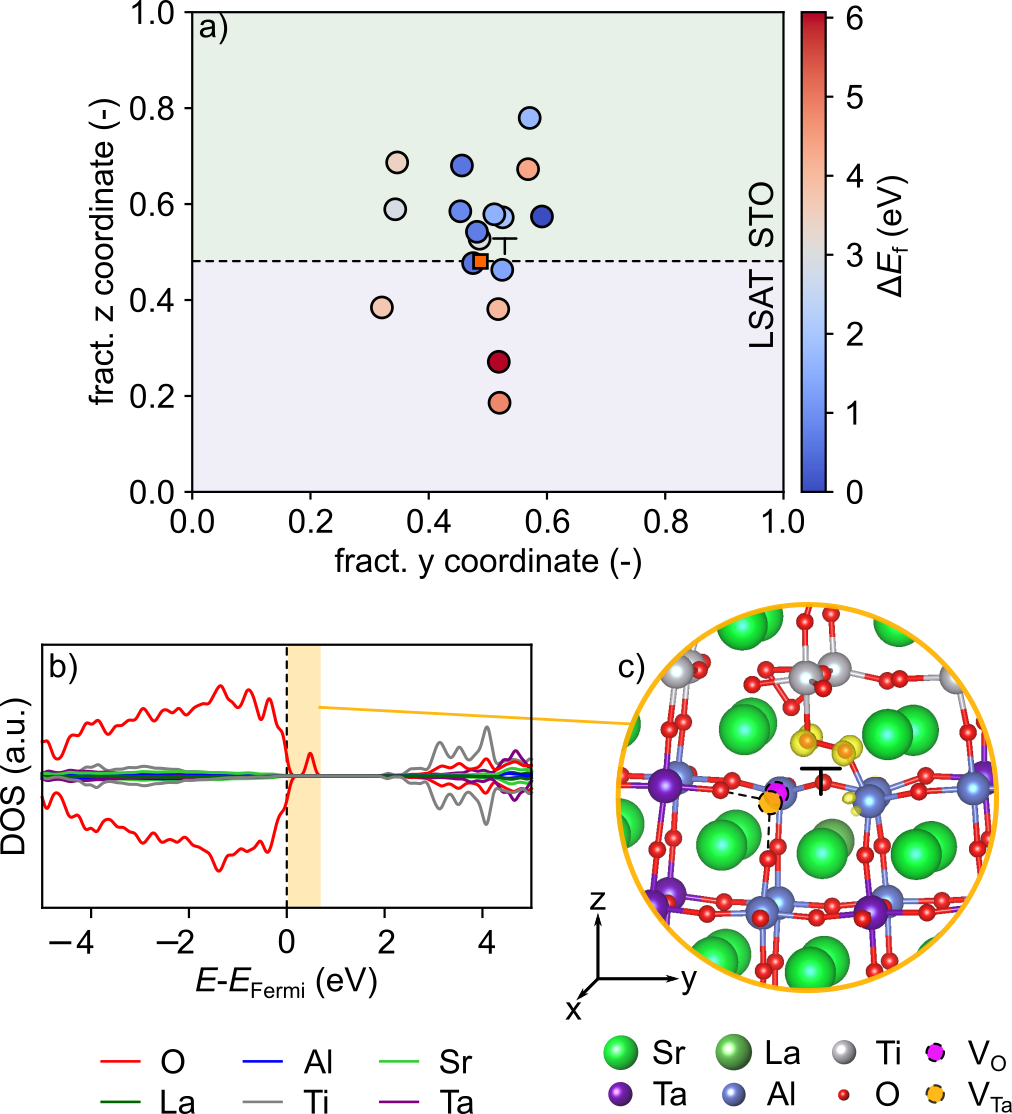}
	\caption{Color map of the relative \ce{V_O} formation energy ($\Delta E_\textrm{f}$) around the dislocation core in presence of a \ce{V_{Ta}}, indicated by the orange square. b) DOS projected on the atoms at the dislocation core containing one \ce{V_{Ta}} at the dislocation core and one \ce{V_O} at a neighboring site. c) charge density isosurfaces ($\unit[10^{-2}]{e/\angstrom^3}$) in the energy range of the defect state highlighted orange in the DOS.}
	\label{fig:VB_BO}
\end{figure}

As expected, the formation of both \ce{V_{Ta}} and \ce{V_{Al}} leads to the creation of holes, as shown by the empty peaks with O character above the Fermi level in Figs.~\ref{fig:VB_pdos}c-e. These states are localized on O atoms at the dislocation core when the vacancy is formed close to the core or, on O atoms adjacent to the defect when the vacancy is further from the core (Figs.~\ref{fig:VB_pdos}f-h). Filling of these defect states at the dislocation core by electron doping, for example, due to oxygen vacancies, could further favor \ce{V_O} formation at the dislocation core and result in conductivity along the dislocation line. For these reasons, we re-evaluated the formation of a \ce{V_O} in presence of a \ce{V_{Ta}}. Comparison of Figures~\ref{fig:VB_BO}a and~\ref{fig:VO_A_pdos}a, suggests that $E_\textrm{f}$ for a \ce{V_O} is lowered due to the \ce{V_{Ta}}, with a reduction of up to 3~eV for O positions close to \ce{V_{Ta}} and both in LSAT and STO. For neighboring \ce{V_{Ta}} and \ce{V_O} at the dislocation core, the excess electrons due to \ce{V_O} formation partially heal the \ce{V_{Ta}} hole state, as can be seen by comparing Figs.~\ref{fig:VB_BO}b and c with Figs.~\ref{fig:VB_pdos}b and e.
%

\subsubsection{Substitutional defects}
%
\begin{figure}
	\centering
	\includegraphics[width=0.8\columnwidth]{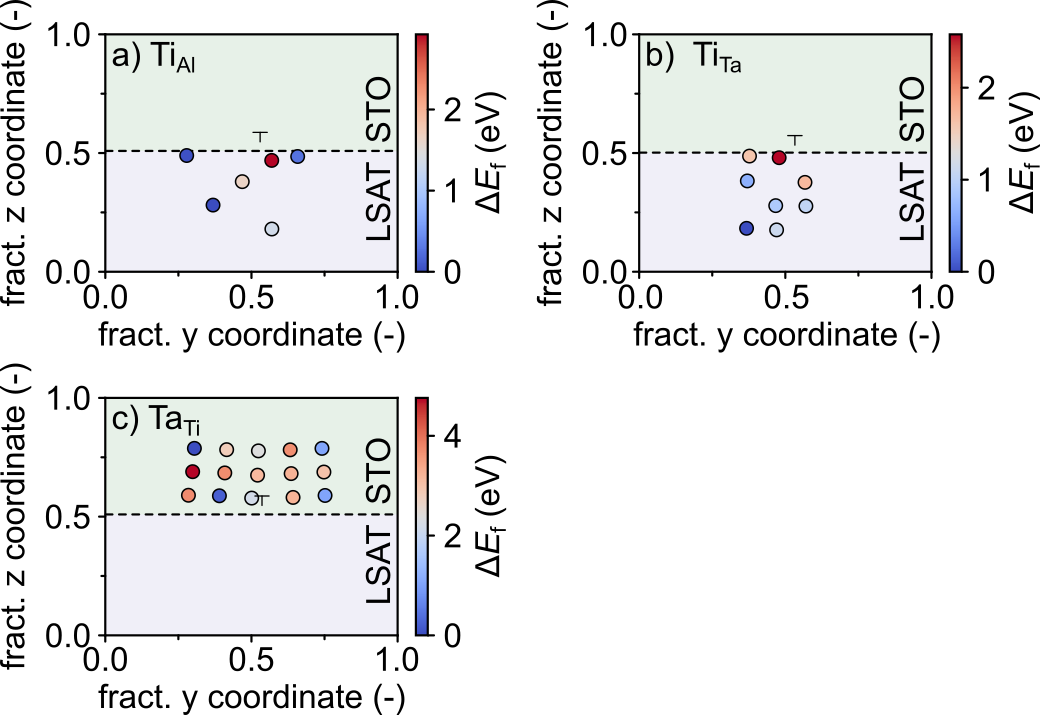}
	\caption{Color map of the relative a) \ce{Ti_{Ta}}, b) \ce{Ti_{Al}}, and c) \ce{Ta_{Ti}} formation energy ($\Delta E_\textrm{f}$) around the dislocation core. The energy of the most stable defect is used in each plot as a reference.}
	\label{fig:sub1}
\end{figure}
Since the interdiffusion of B-site cations, namely Ti, Al, and Ta, between LSAT and STO is clearly visible in the EDX maps (Fig.~\ref{fig:edx}), we also considered the possibility of Ti substituting Al (\ce{Ti_{Al}}) or Ta (\ce{Ti_{Ta}}) in the LSAT substrate and of Ta substituting Ti (\ce{Ta_{Ti}}) in the STO film. When \ce{Ta_{Ti}} are formed in STO (Fig.~\ref{fig:sub1}c), we observe fairly strong variations in the formation energy for the explored configurations, with low values at the interface adjacent to the dislocation core, but also further up in STO, in line with EDX maps showing Ta diffusion to 3 or even more Ti layers from the interface. Results for Ti diffusion into LSAT, forming \ce{Ti_{Ta}} or \ce{Ti_{Al}} (Figs.~\ref{fig:sub1}a and b) are, instead, at odd with experiments, and suggest that Ti should not be mainly located at the dislocation core, contrarily to what is shown in Fig.~\ref{fig:edx}c).

\begin{figure}
	\centering
	\includegraphics[width=0.8\columnwidth]{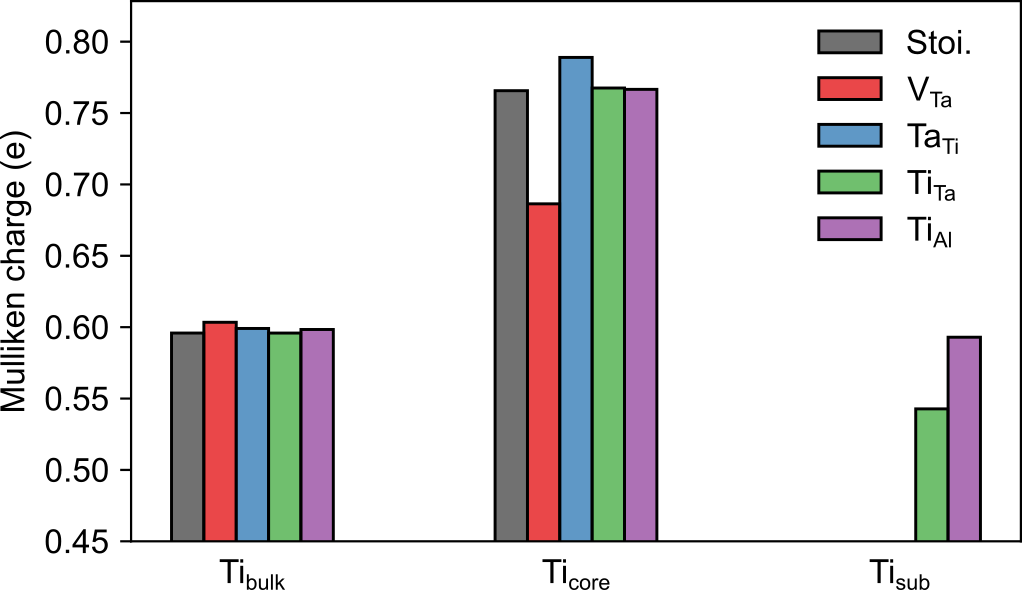}
	\caption{Average Mulliken charges for a) Ti atoms in STO far from the dislocation core (Ti$_\mathrm{bulk}$), b) Ti atoms close to the dislocation core (Ti$_\mathrm{core}$), and c) for  substitutional Ti atoms in LSAT (Ti$_\mathrm{sub}$).}
	\label{fig:defchg}
\end{figure}

\begin{figure}
	\centering
	\includegraphics[width=0.8\columnwidth]{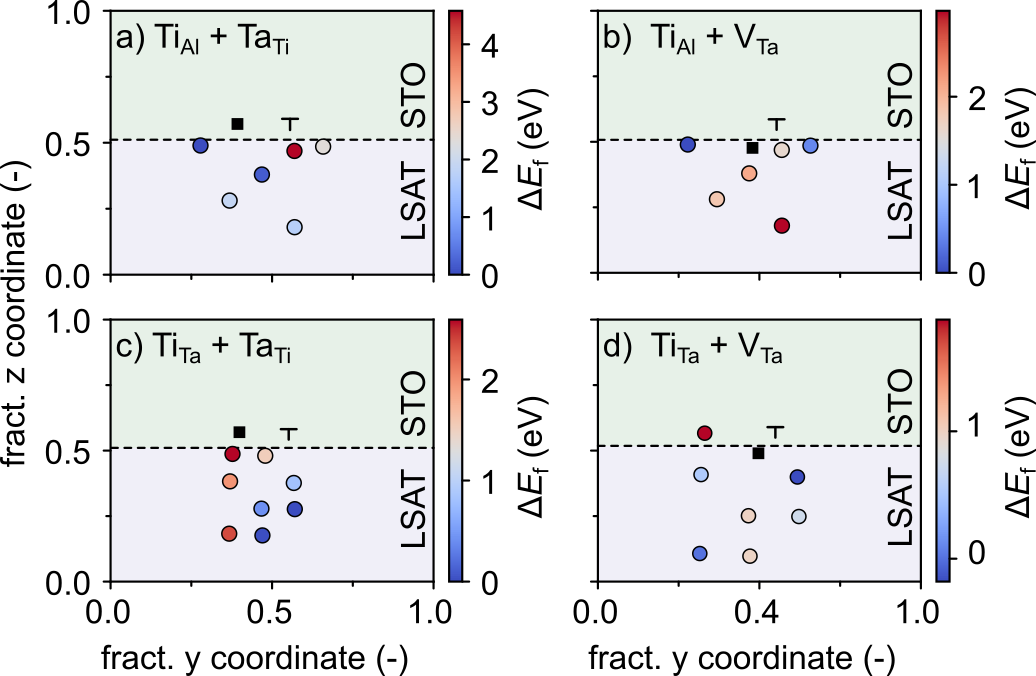}
	\caption{Color map of the relative formation energy ($\Delta E_\textrm{f}$) of a \ce{Ti_{Al}} in presence of one a) \ce{Ta_{Ti}} or b) \ce{V_{Ta}} and of a \ce{Ti_{Ta}} in presence of one c) \ce{Ta_{Ti}} or d) \ce{V_{Ta}}. The energy of the most stable defect is used in each plot as a reference. The black square indicates the position of the \ce{Ta_{Ti}} or \ce{V_{Ta}} defect.}
	\label{fig:sub2}
\end{figure}

This result suggests that the Ti/Ta/Al intermixing observed experimentally could be the result of a complex interplay between different defect types, some of which can lead to excess charges. To further investigate the effect of electron or hole doping on the cation interdiffusion, we select the two example defects \ce{Ta_{Ti}} and \ce{V_{Ta}} that result in electron and hole doping respectively, and reevaluate \ce{Ti_{Al}} and \ce{Ti_{Ta}} formation in LSAT in presence of these defects. Fig.~\ref{fig:defchg}, reports the average Mulliken charges for Ti atoms in STO (Ti$_\mathrm{bulk}$) or at the dislocation core (Ti$_{core}$) for different single defects in the dislocation model. This data shows, indeed, that while Ti$_\mathrm{bulk}$ stays unaltered, Ti$_\mathrm{core}$ are oxidized/reduced in presence of a \ce{Ta_{Ti}}/\ce{V_{Ta}} compared to the stoichiometric model. Instead, Ti substitution in LSAT (\ce{Ti_{Ta}} or \ce{Ti_{Al}}) does not significantly alter charges of Ti$_\mathrm{bulk}$ or Ti$_\mathrm{core}$, but instead the substitutional Ti atom (Ti$_\mathrm{sub}$) is oxidized compared to Ti$_\mathrm{bulk}$, especially when substitution takes place at the Ta site. Comparison of Fig.~\ref{fig:sub1} and Fig.~\ref{fig:sub2} indicates that reducing \ce{Ta_{Ti}} leading to electron-doping at the dislocation core favors Ti substitution around the dislocation at less oxidizing \ce{Al^{3+}} sites, while hole-doping by \ce{V_{Ta}} at the dislocation core increases Ti substitution at the more oxidizing \ce{Ta^{5+}} site, due to charge compensation reasons. In summary, this data shows that while Ti diffusion into LSAT is unlikely without other defects, it is facilitated by the simultaneous presence of \ce{V_{Ta}} or by Ta diffusion into STO (\ce{Ta_{Ti}}).
%

\section{Conclusions}

In this work, we investigated the atomic structure and defect chemistry of the highly ordered Moir\'e network of dislocations formed at the interface between an STO thin film grown on a LSAT substrate, combining experimental and theoretical techniques. Combined MD and DFT calculations lead to an atomic scale model of the dislocation core that features undercoordinated Ta/Al cations at the interface and has a strain field in nice agreement with the one derived from high-resolution STEM data.

Both EELS and DFT results indicate that oxygen vacancies (\ce{V_O}) easily form at the dislocation core and in STO above the dislocation core. Furthermore, DFT calculations show that the experimentally observed \ce{Ti^{3+}} around the dislocation core are due to both the dislocation structure itself as well as the presence of \ce{V_O}.

EDX mapping suggests that cation vacancies form at the dislocation core in LSAT: Ti substitutes Al and Ta around the dislocation core in LSAT and a partial substitution of Ta by Ti takes place in the STO film above the dislocation core. DFT calculations confirm that cation vacancies are favored to form in a compressively strained region below the dislocation core, leading to hole doping that, in turn, further favors \ce{V_O} formation at the core. Finally, we show that Ti diffusion into the LSAT substrate below the dislocation core only occurs in presence of cation vacancies (favoring substitutions at Ta sites) or concurrently with the diffusion of Ta into STO (favoring substitution at the Al site).

Even though additional defect combinations and sites further from the dislocation core could be explored, the present DFT results and their good agreement with experiment lead to a deeper understanding of the structure and electronic properties of these systems. Our results, show, in particular, the predominance of p-type 1D conductivity along the dislocations, depending on the defects present also with shallow acceptor states. These results will be instrumental in further engineering the functional properties of these systems.

\section*{Acknowledgements}
	CS and ES were supported by the NCCR MARVEL, a National Centre of Competence in Research, funded by the Swiss National Science Foundation (grant number 182892). ES was also supported by the European Union’s Horizon 2020 research and innovation program under the Marie Sklodowska-Curie grant agreement No 884104 (PSI-FELLOW-III-3i).  Computational resources were provided by the University of Bern (on the HPC cluster UBELIX, http://www.id.unibe.ch/hpc) and by the Swiss National Supercomputing Center (CSCS) under project ID mr26. 

\bibliography{references}

\begin{thebibliography}{39}%
\makeatletter
\providecommand \@ifxundefined [1]{%
 \@ifx{#1\undefined}
}%
\providecommand \@ifnum [1]{%
 \ifnum #1\expandafter \@firstoftwo
 \else \expandafter \@secondoftwo
 \fi
}%
\providecommand \@ifx [1]{%
 \ifx #1\expandafter \@firstoftwo
 \else \expandafter \@secondoftwo
 \fi
}%
\providecommand \natexlab [1]{#1}%
\providecommand \enquote  [1]{``#1''}%
\providecommand \bibnamefont  [1]{#1}%
\providecommand \bibfnamefont [1]{#1}%
\providecommand \citenamefont [1]{#1}%
\providecommand \href@noop [0]{\@secondoftwo}%
\providecommand \href [0]{\begingroup \@sanitize@url \@href}%
\providecommand \@href[1]{\@@startlink{#1}\@@href}%
\providecommand \@@href[1]{\endgroup#1\@@endlink}%
\providecommand \@sanitize@url [0]{\catcode `\\12\catcode `\$12\catcode `\&12\catcode `\#12\catcode `\^12\catcode `\_12\catcode `\%12\relax}%
\providecommand \@@startlink[1]{}%
\providecommand \@@endlink[0]{}%
\providecommand \url  [0]{\begingroup\@sanitize@url \@url }%
\providecommand \@url [1]{\endgroup\@href {#1}{\urlprefix }}%
\providecommand \urlprefix  [0]{URL }%
\providecommand \Eprint [0]{\href }%
\providecommand \doibase [0]{http://dx.doi.org/}%
\providecommand \selectlanguage [0]{\@gobble}%
\providecommand \bibinfo  [0]{\@secondoftwo}%
\providecommand \bibfield  [0]{\@secondoftwo}%
\providecommand \translation [1]{[#1]}%
\providecommand \BibitemOpen [0]{}%
\providecommand \bibitemStop [0]{}%
\providecommand \bibitemNoStop [0]{.\EOS\space}%
\providecommand \EOS [0]{\spacefactor3000\relax}%
\providecommand \BibitemShut  [1]{\csname bibitem#1\endcsname}%
\let\auto@bib@innerbib\@empty
\bibitem [{\citenamefont {Zubko}\ \emph {et~al.}(2011)\citenamefont {Zubko}, \citenamefont {Gariglio}, \citenamefont {Gabay}, \citenamefont {Ghosez},\ and\ \citenamefont {Triscone}}]{Zubko2011}%
  \BibitemOpen
  \bibfield  {author} {\bibinfo {author} {\bibfnamefont {P.}~\bibnamefont {Zubko}}, \bibinfo {author} {\bibfnamefont {S.}~\bibnamefont {Gariglio}}, \bibinfo {author} {\bibfnamefont {M.}~\bibnamefont {Gabay}}, \bibinfo {author} {\bibfnamefont {Ph.}\ \bibnamefont {Ghosez}}, \ and\ \bibinfo {author} {\bibfnamefont {J.-M.}\ \bibnamefont {Triscone}},\ }\bibfield  {title} {\enquote {\bibinfo {title} {{I}nterface physics in complex oxide heterostructures},}\ }\href {\doibase 10.1146/annurev-conmatphys-062910-140445} {\bibfield  {journal} {\bibinfo  {journal} {Annu. Rev. Condens. Matter Phys.}\ }\textbf {\bibinfo {volume} {2}},\ \bibinfo {pages} {141--165} (\bibinfo {year} {2011})}\BibitemShut {NoStop}%
\bibitem [{\citenamefont {Lin}\ \emph {et~al.}(2013)\citenamefont {Lin}, \citenamefont {Zhu}, \citenamefont {Fauqu{\'e}},\ and\ \citenamefont {Behnia}}]{Lin2013}%
  \BibitemOpen
  \bibfield  {author} {\bibinfo {author} {\bibfnamefont {X.}~\bibnamefont {Lin}}, \bibinfo {author} {\bibfnamefont {Z.}~\bibnamefont {Zhu}}, \bibinfo {author} {\bibfnamefont {B.}~\bibnamefont {Fauqu{\'e}}}, \ and\ \bibinfo {author} {\bibfnamefont {K.}~\bibnamefont {Behnia}},\ }\bibfield  {title} {\enquote {\bibinfo {title} {{F}ermi surface of the most dilute superconductor},}\ }\href {\doibase 10.1103/PhysRevX.3.021002} {\bibfield  {journal} {\bibinfo  {journal} {Phys. Rev. X}\ }\textbf {\bibinfo {volume} {3}},\ \bibinfo {pages} {021002} (\bibinfo {year} {2013})}\BibitemShut {NoStop}%
\bibitem [{\citenamefont {Bhattacharya}\ and\ \citenamefont {May}(2014)}]{Bhattacharya2014}%
  \BibitemOpen
  \bibfield  {author} {\bibinfo {author} {\bibfnamefont {A.}~\bibnamefont {Bhattacharya}}\ and\ \bibinfo {author} {\bibfnamefont {S.~J.}\ \bibnamefont {May}},\ }\bibfield  {title} {\enquote {\bibinfo {title} {{M}agnetic oxide heterostructures},}\ }\href {\doibase 10.1146/annurev-matsci-070813-113447} {\bibfield  {journal} {\bibinfo  {journal} {Annu. Rev. Mater. Res.}\ }\textbf {\bibinfo {volume} {44}},\ \bibinfo {pages} {65--90} (\bibinfo {year} {2014})}\BibitemShut {NoStop}%
\bibitem [{\citenamefont {Arandiyan}\ \emph {et~al.}(2021)\citenamefont {Arandiyan}, \citenamefont {Mofarah}, \citenamefont {Sorrell}, \citenamefont {Doustkhah}, \citenamefont {Sajjadi}, \citenamefont {Hao}, \citenamefont {Wang}, \citenamefont {Sun}, \citenamefont {Ni}, \citenamefont {Rezaei}, \citenamefont {Shao},\ and\ \citenamefont {Maschmeyer}}]{Arandiyan2021}%
  \BibitemOpen
  \bibfield  {author} {\bibinfo {author} {\bibfnamefont {H.}~\bibnamefont {Arandiyan}}, \bibinfo {author} {\bibfnamefont {S.~S.}\ \bibnamefont {Mofarah}}, \bibinfo {author} {\bibfnamefont {C.~C.}\ \bibnamefont {Sorrell}}, \bibinfo {author} {\bibfnamefont {E.}~\bibnamefont {Doustkhah}}, \bibinfo {author} {\bibfnamefont {B.}~\bibnamefont {Sajjadi}}, \bibinfo {author} {\bibfnamefont {D.}~\bibnamefont {Hao}}, \bibinfo {author} {\bibfnamefont {Y.}~\bibnamefont {Wang}}, \bibinfo {author} {\bibfnamefont {H.}~\bibnamefont {Sun}}, \bibinfo {author} {\bibfnamefont {B.-J.}\ \bibnamefont {Ni}}, \bibinfo {author} {\bibfnamefont {M.}~\bibnamefont {Rezaei}}, \bibinfo {author} {\bibfnamefont {Z.}~\bibnamefont {Shao}}, \ and\ \bibinfo {author} {\bibfnamefont {Th.}\ \bibnamefont {Maschmeyer}},\ }\bibfield  {title} {\enquote {\bibinfo {title} {{D}efect engineering of oxide perovskites for catalysis and energy storage: synthesis of chemistry and materials science},}\ }\href {\doibase 10.1039/D0CS00639D} {\bibfield  {journal} {\bibinfo  {journal} {Chem. Soc. Rev.}\ }\textbf {\bibinfo {volume} {50}},\ \bibinfo {pages} {10116--10211} (\bibinfo {year} {2021})}\BibitemShut {NoStop}%
\bibitem [{\citenamefont {Kalinin}\ and\ \citenamefont {Spaldin}(2013)}]{Kalinin2013}%
  \BibitemOpen
  \bibfield  {author} {\bibinfo {author} {\bibfnamefont {S.~V.}\ \bibnamefont {Kalinin}}\ and\ \bibinfo {author} {\bibfnamefont {N.~A.}\ \bibnamefont {Spaldin}},\ }\bibfield  {title} {\enquote {\bibinfo {title} {{F}unctional ion defects in transition metal oxides},}\ }\href {\doibase 10.1126/science.1243098} {\bibfield  {journal} {\bibinfo  {journal} {Science}\ }\textbf {\bibinfo {volume} {341}},\ \bibinfo {pages} {858--859} (\bibinfo {year} {2013})}\BibitemShut {NoStop}%
\bibitem [{\citenamefont {Chandrasena}\ \emph {et~al.}(2017)\citenamefont {Chandrasena}, \citenamefont {Yang}, \citenamefont {Lei}, \citenamefont {Delgado-Jaime}, \citenamefont {Wijesekara}, \citenamefont {Golalikhani}, \citenamefont {Davidson}, \citenamefont {Arenholz}, \citenamefont {Kobayashi}, \citenamefont {Kobata}, \citenamefont {de~Groot}, \citenamefont {Aschauer}, \citenamefont {Spaldin}, \citenamefont {Xi},\ and\ \citenamefont {Gray}}]{Chandrasena2017}%
  \BibitemOpen
  \bibfield  {author} {\bibinfo {author} {\bibfnamefont {R.~U.}\ \bibnamefont {Chandrasena}}, \bibinfo {author} {\bibfnamefont {W.}~\bibnamefont {Yang}}, \bibinfo {author} {\bibfnamefont {Q.}~\bibnamefont {Lei}}, \bibinfo {author} {\bibfnamefont {M.~U.}\ \bibnamefont {Delgado-Jaime}}, \bibinfo {author} {\bibfnamefont {K.~D.}\ \bibnamefont {Wijesekara}}, \bibinfo {author} {\bibfnamefont {M.}~\bibnamefont {Golalikhani}}, \bibinfo {author} {\bibfnamefont {B.~A.}\ \bibnamefont {Davidson}}, \bibinfo {author} {\bibfnamefont {E.}~\bibnamefont {Arenholz}}, \bibinfo {author} {\bibfnamefont {K.}~\bibnamefont {Kobayashi}}, \bibinfo {author} {\bibfnamefont {M.}~\bibnamefont {Kobata}}, \bibinfo {author} {\bibfnamefont {F.~M.~F.}\ \bibnamefont {de~Groot}}, \bibinfo {author} {\bibfnamefont {U.}~\bibnamefont {Aschauer}}, \bibinfo {author} {\bibfnamefont {N.~A.}\ \bibnamefont {Spaldin}}, \bibinfo {author} {\bibfnamefont {X.}~\bibnamefont {Xi}}, \ and\ \bibinfo {author} {\bibfnamefont {A.~X.}\ \bibnamefont {Gray}},\ }\bibfield  {title} {\enquote {\bibinfo {title} {{S}train-engineered oxygen vacancies in \ce{CaMnO3} thin films},}\ }\href {\doibase 10.1021/acs.nanolett.6b03986} {\bibfield  {journal} {\bibinfo  {journal} {Nano Lett.}\ }\textbf {\bibinfo {volume} {17}},\ \bibinfo {pages} {794--799} (\bibinfo {year} {2017})}\BibitemShut {NoStop}%
\bibitem [{\citenamefont {Ricca}\ \emph {et~al.}(2020{\natexlab{a}})\citenamefont {Ricca}, \citenamefont {Niederhauser},\ and\ \citenamefont {Aschauer}}]{Ricca2020}%
  \BibitemOpen
  \bibfield  {author} {\bibinfo {author} {\bibfnamefont {C.}~\bibnamefont {Ricca}}, \bibinfo {author} {\bibfnamefont {N.}~\bibnamefont {Niederhauser}}, \ and\ \bibinfo {author} {\bibfnamefont {U.}~\bibnamefont {Aschauer}},\ }\bibfield  {title} {\enquote {\bibinfo {title} {{L}ocal polarization in oxygen-deficient \ce{LaMnO3} induced by charge localization in the {J}ahn-{T}eller distorted structure},}\ }\href {\doibase 10.1103/PhysRevResearch.2.042040} {\bibfield  {journal} {\bibinfo  {journal} {Phys. Rev. Res.}\ }\textbf {\bibinfo {volume} {2}},\ \bibinfo {pages} {042040} (\bibinfo {year} {2020}{\natexlab{a}})}\BibitemShut {NoStop}%
\bibitem [{\citenamefont {Ohta}\ \emph {et~al.}(2007)\citenamefont {Ohta}, \citenamefont {Kim}, \citenamefont {Mune}, \citenamefont {Mizoguchi}, \citenamefont {Nomura}, \citenamefont {Ohta}, \citenamefont {Nomura}, \citenamefont {Nakanishi}, \citenamefont {Ikuhara}, \citenamefont {Hirano}, \citenamefont {Hosono},\ and\ \citenamefont {Koumoto}}]{Ohta2007}%
  \BibitemOpen
  \bibfield  {author} {\bibinfo {author} {\bibfnamefont {H.}~\bibnamefont {Ohta}}, \bibinfo {author} {\bibfnamefont {S.}~\bibnamefont {Kim}}, \bibinfo {author} {\bibfnamefont {Y.}~\bibnamefont {Mune}}, \bibinfo {author} {\bibfnamefont {T.}~\bibnamefont {Mizoguchi}}, \bibinfo {author} {\bibfnamefont {K.}~\bibnamefont {Nomura}}, \bibinfo {author} {\bibfnamefont {S.}~\bibnamefont {Ohta}}, \bibinfo {author} {\bibfnamefont {T.}~\bibnamefont {Nomura}}, \bibinfo {author} {\bibfnamefont {Y.}~\bibnamefont {Nakanishi}}, \bibinfo {author} {\bibfnamefont {Y.}~\bibnamefont {Ikuhara}}, \bibinfo {author} {\bibfnamefont {M.}~\bibnamefont {Hirano}}, \bibinfo {author} {\bibfnamefont {H.}~\bibnamefont {Hosono}}, \ and\ \bibinfo {author} {\bibfnamefont {K.}~\bibnamefont {Koumoto}},\ }\bibfield  {title} {\enquote {\bibinfo {title} {{G}iant thermoelectric {S}eebeck coefficient of a two-dimensional electron gas in \ce{SrTiO3}},}\ }\href {\doibase 10.1038/nmat1821} {\bibfield  {journal} {\bibinfo  {journal} {Nat. Mater.}\ }\textbf {\bibinfo {volume} {6}},\ \bibinfo {pages} {129--134} (\bibinfo {year} {2007})}\BibitemShut {NoStop}%
\bibitem [{\citenamefont {Garcia-Barriocanal}\ \emph {et~al.}(2008)\citenamefont {Garcia-Barriocanal}, \citenamefont {Rivera-Calzada}, \citenamefont {Varela}, \citenamefont {Sefrioui}, \citenamefont {Iborra}, \citenamefont {Leon}, \citenamefont {Pennycook},\ and\ \citenamefont {Santamaria}}]{Barriocanal2008}%
  \BibitemOpen
  \bibfield  {author} {\bibinfo {author} {\bibfnamefont {J.}~\bibnamefont {Garcia-Barriocanal}}, \bibinfo {author} {\bibfnamefont {A.}~\bibnamefont {Rivera-Calzada}}, \bibinfo {author} {\bibfnamefont {M.}~\bibnamefont {Varela}}, \bibinfo {author} {\bibfnamefont {Z.}~\bibnamefont {Sefrioui}}, \bibinfo {author} {\bibfnamefont {E.}~\bibnamefont {Iborra}}, \bibinfo {author} {\bibfnamefont {C.}~\bibnamefont {Leon}}, \bibinfo {author} {\bibfnamefont {S.~J.}\ \bibnamefont {Pennycook}}, \ and\ \bibinfo {author} {\bibfnamefont {J.}~\bibnamefont {Santamaria}},\ }\bibfield  {title} {\enquote {\bibinfo {title} {{C}olossal ionic conductivity at interfaces of epitaxial \ce{ZrO2}:\ce{Y2O3}/\ce{SrTiO3} heterostructures},}\ }\href {\doibase 10.1126/science.1156393} {\bibfield  {journal} {\bibinfo  {journal} {Science}\ }\textbf {\bibinfo {volume} {321}},\ \bibinfo {pages} {676--680} (\bibinfo {year} {2008})}\BibitemShut {NoStop}%
\bibitem [{\citenamefont {Mannhart}\ and\ \citenamefont {Schlom}(2010)}]{Mannhart2010}%
  \BibitemOpen
  \bibfield  {author} {\bibinfo {author} {\bibfnamefont {J.}~\bibnamefont {Mannhart}}\ and\ \bibinfo {author} {\bibfnamefont {D.~G.}\ \bibnamefont {Schlom}},\ }\bibfield  {title} {\enquote {\bibinfo {title} {{O}xide interfaces-{A}n opportunity for electronics},}\ }\href {\doibase 10.1126/science.1181862} {\bibfield  {journal} {\bibinfo  {journal} {Science}\ }\textbf {\bibinfo {volume} {327}},\ \bibinfo {pages} {1607--1611} (\bibinfo {year} {2010})}\BibitemShut {NoStop}%
\bibitem [{\citenamefont {Zubko}\ \emph {et~al.}(2012)\citenamefont {Zubko}, \citenamefont {Jecklin}, \citenamefont {Torres-Pardo}, \citenamefont {Aguado-Puente}, \citenamefont {Gloter}, \citenamefont {Lichtensteiger}, \citenamefont {Junquera}, \citenamefont {St{\'e}phan},\ and\ \citenamefont {Triscone}}]{Zubko2012}%
  \BibitemOpen
  \bibfield  {author} {\bibinfo {author} {\bibfnamefont {P.}~\bibnamefont {Zubko}}, \bibinfo {author} {\bibfnamefont {N.}~\bibnamefont {Jecklin}}, \bibinfo {author} {\bibfnamefont {A.}~\bibnamefont {Torres-Pardo}}, \bibinfo {author} {\bibfnamefont {P.}~\bibnamefont {Aguado-Puente}}, \bibinfo {author} {\bibfnamefont {A.}~\bibnamefont {Gloter}}, \bibinfo {author} {\bibfnamefont {C.}~\bibnamefont {Lichtensteiger}}, \bibinfo {author} {\bibfnamefont {J.}~\bibnamefont {Junquera}}, \bibinfo {author} {\bibfnamefont {O.}~\bibnamefont {St{\'e}phan}}, \ and\ \bibinfo {author} {\bibfnamefont {J.-M.}\ \bibnamefont {Triscone}},\ }\bibfield  {title} {\enquote {\bibinfo {title} {{E}lectrostatic coupling and local structural distortions at interfaces in ferroelectric/paraelectric superlattices},}\ }\href {\doibase 10.1021/nl3003717} {\bibfield  {journal} {\bibinfo  {journal} {Nano Lett.}\ }\textbf {\bibinfo {volume} {12}},\ \bibinfo {pages} {2846--2851} (\bibinfo {year} {2012})}\BibitemShut {NoStop}%
\bibitem [{\citenamefont {Chen}\ \emph {et~al.}(2017)\citenamefont {Chen}, \citenamefont {Green}, \citenamefont {Sutarto}, \citenamefont {He}, \citenamefont {Linderoth}, \citenamefont {Sawatzky},\ and\ \citenamefont {Pryds}}]{Chen2017}%
  \BibitemOpen
  \bibfield  {author} {\bibinfo {author} {\bibfnamefont {Y.}~\bibnamefont {Chen}}, \bibinfo {author} {\bibfnamefont {Robert~J.}\ \bibnamefont {Green}}, \bibinfo {author} {\bibfnamefont {R.}~\bibnamefont {Sutarto}}, \bibinfo {author} {\bibfnamefont {F.}~\bibnamefont {He}}, \bibinfo {author} {\bibfnamefont {S.}~\bibnamefont {Linderoth}}, \bibinfo {author} {\bibfnamefont {G.~A.}\ \bibnamefont {Sawatzky}}, \ and\ \bibinfo {author} {\bibfnamefont {N.}~\bibnamefont {Pryds}},\ }\bibfield  {title} {\enquote {\bibinfo {title} {{T}uning the two-dimensional electron liquid at oxide interfaces by buffer-layer-engineered redox reactions},}\ }\href {\doibase 10.1021/acs.nanolett.7b03744} {\bibfield  {journal} {\bibinfo  {journal} {Nano Lett.}\ }\textbf {\bibinfo {volume} {17}},\ \bibinfo {pages} {7062--7066} (\bibinfo {year} {2017})}\BibitemShut {NoStop}%
\bibitem [{\citenamefont {Burian}\ \emph {et~al.}(2021)\citenamefont {Burian}, \citenamefont {Pedrini}, \citenamefont {Ortiz~Hernandez}, \citenamefont {Ueda}, \citenamefont {Vaz}, \citenamefont {Caputo}, \citenamefont {Radovic},\ and\ \citenamefont {Staub}}]{Burian2021}%
  \BibitemOpen
  \bibfield  {author} {\bibinfo {author} {\bibfnamefont {M.}~\bibnamefont {Burian}}, \bibinfo {author} {\bibfnamefont {B.~F.}\ \bibnamefont {Pedrini}}, \bibinfo {author} {\bibfnamefont {N.}~\bibnamefont {Ortiz~Hernandez}}, \bibinfo {author} {\bibfnamefont {H.}~\bibnamefont {Ueda}}, \bibinfo {author} {\bibfnamefont {C.~A.~F.}\ \bibnamefont {Vaz}}, \bibinfo {author} {\bibfnamefont {M.}~\bibnamefont {Caputo}}, \bibinfo {author} {\bibfnamefont {M.}~\bibnamefont {Radovic}}, \ and\ \bibinfo {author} {\bibfnamefont {U.}~\bibnamefont {Staub}},\ }\bibfield  {title} {\enquote {\bibinfo {title} {{B}uried moir{\'e} supercells through \ce{SrTiO3} nanolayer relaxation},}\ }\href {\doibase 10.1103/PhysRevResearch.3.013225} {\bibfield  {journal} {\bibinfo  {journal} {Phys. Rev. Res.}\ }\textbf {\bibinfo {volume} {3}},\ \bibinfo {pages} {013225} (\bibinfo {year} {2021})}\BibitemShut {NoStop}%
\bibitem [{\citenamefont {Hirel}\ \emph {et~al.}(2012)\citenamefont {Hirel}, \citenamefont {Mrovec},\ and\ \citenamefont {Els{\"a}sser}}]{HIREL2012329}%
  \BibitemOpen
  \bibfield  {author} {\bibinfo {author} {\bibfnamefont {P.}~\bibnamefont {Hirel}}, \bibinfo {author} {\bibfnamefont {M.}~\bibnamefont {Mrovec}}, \ and\ \bibinfo {author} {\bibfnamefont {C.}~\bibnamefont {Els{\"a}sser}},\ }\bibfield  {title} {\enquote {\bibinfo {title} {{A}tomistic simulation study of (110) dislocations in strontium titanate},}\ }\href {\doibase 10.1016/j.actamat.2011.09.049} {\bibfield  {journal} {\bibinfo  {journal} {Acta Mater.}\ }\textbf {\bibinfo {volume} {60}},\ \bibinfo {pages} {329--338} (\bibinfo {year} {2012})}\BibitemShut {NoStop}%
\bibitem [{\citenamefont {Sun}\ \emph {et~al.}(2015)\citenamefont {Sun}, \citenamefont {Marrocchelli},\ and\ \citenamefont {Yildiz}}]{sun2015}%
  \BibitemOpen
  \bibfield  {author} {\bibinfo {author} {\bibfnamefont {L.}~\bibnamefont {Sun}}, \bibinfo {author} {\bibfnamefont {D.}~\bibnamefont {Marrocchelli}}, \ and\ \bibinfo {author} {\bibfnamefont {B.}~\bibnamefont {Yildiz}},\ }\bibfield  {title} {\enquote {\bibinfo {title} {{E}dge dislocation slows down oxide ion diffusion in doped \ce{CeO2} by segregation of charged defects},}\ }\href {\doibase 10.1038/ncomms7294} {\bibfield  {journal} {\bibinfo  {journal} {Nat. Commun.}\ }\textbf {\bibinfo {volume} {6}},\ \bibinfo {pages} {1--10} (\bibinfo {year} {2015})}\BibitemShut {NoStop}%
\bibitem [{\citenamefont {Marrocchelli}\ \emph {et~al.}(2015)\citenamefont {Marrocchelli}, \citenamefont {Sun},\ and\ \citenamefont {Yildiz}}]{Marrocchelli2015}%
  \BibitemOpen
  \bibfield  {author} {\bibinfo {author} {\bibfnamefont {D.}~\bibnamefont {Marrocchelli}}, \bibinfo {author} {\bibfnamefont {L.}~\bibnamefont {Sun}}, \ and\ \bibinfo {author} {\bibfnamefont {B.}~\bibnamefont {Yildiz}},\ }\bibfield  {title} {\enquote {\bibinfo {title} {{D}islocations in \ce{SrTiO3}: {E}asy to reduce but not so fast for oxygen transport},}\ }\href {\doibase 10.1021/ja513176u} {\bibfield  {journal} {\bibinfo  {journal} {J. Am. Chem. Soc.}\ }\textbf {\bibinfo {volume} {137}},\ \bibinfo {pages} {4735--4748} (\bibinfo {year} {2015})}\BibitemShut {NoStop}%
\bibitem [{\citenamefont {Plimpton}(1995)}]{PLIMPTON19951}%
  \BibitemOpen
  \bibfield  {author} {\bibinfo {author} {\bibfnamefont {S.}~\bibnamefont {Plimpton}},\ }\bibfield  {title} {\enquote {\bibinfo {title} {{F}ast parallel algorithms for short-range molecular dynamics},}\ }\href {\doibase 10.1006/jcph.1995.1039} {\bibfield  {journal} {\bibinfo  {journal} {J. Comp. Phys.}\ }\textbf {\bibinfo {volume} {117}},\ \bibinfo {pages} {1--19} (\bibinfo {year} {1995})}\BibitemShut {NoStop}%
\bibitem [{\citenamefont {Lewis}\ and\ \citenamefont {Catlow}(1985)}]{Lewis_1985}%
  \BibitemOpen
  \bibfield  {author} {\bibinfo {author} {\bibfnamefont {G.~V.}\ \bibnamefont {Lewis}}\ and\ \bibinfo {author} {\bibfnamefont {C.~R.~A.}\ \bibnamefont {Catlow}},\ }\bibfield  {title} {\enquote {\bibinfo {title} {{P}otential models for ionic oxides},}\ }\href {\doibase 10.1088/0022-3719/18/6/010} {\bibfield  {journal} {\bibinfo  {journal} {J. Phys. C: Solid State Phys.}\ }\textbf {\bibinfo {volume} {18}},\ \bibinfo {pages} {1149} (\bibinfo {year} {1985})}\BibitemShut {NoStop}%
\bibitem [{\citenamefont {Gale}(1997)}]{gulp1}%
  \BibitemOpen
  \bibfield  {author} {\bibinfo {author} {\bibfnamefont {J.~D.}\ \bibnamefont {Gale}},\ }\bibfield  {title} {\enquote {\bibinfo {title} {{GULP}: {A} computer program for the symmetry-adapted simulation of solids},}\ }\href {\doibase 10.1039/A606455H} {\bibfield  {journal} {\bibinfo  {journal} {J. Chem. Soc., Faraday Trans.}\ }\textbf {\bibinfo {volume} {93}},\ \bibinfo {pages} {629--637} (\bibinfo {year} {1997})}\BibitemShut {NoStop}%
\bibitem [{\citenamefont {Gale}\ and\ \citenamefont {Rohl}(2003)}]{gulp2}%
  \BibitemOpen
  \bibfield  {author} {\bibinfo {author} {\bibfnamefont {J.~D.}\ \bibnamefont {Gale}}\ and\ \bibinfo {author} {\bibfnamefont {A.~L.}\ \bibnamefont {Rohl}},\ }\bibfield  {title} {\enquote {\bibinfo {title} {{T}he general utility lattice program ({GULP})},}\ }\href {\doibase 10.1080/0892702031000104887} {\bibfield  {journal} {\bibinfo  {journal} {Mol. Simul.}\ }\textbf {\bibinfo {volume} {29}},\ \bibinfo {pages} {291--341} (\bibinfo {year} {2003})}\BibitemShut {NoStop}%
\bibitem [{\citenamefont {Gale}(2005)}]{gulp3}%
  \BibitemOpen
  \bibfield  {author} {\bibinfo {author} {\bibfnamefont {J.~D.}\ \bibnamefont {Gale}},\ }\bibfield  {title} {\enquote {\bibinfo {title} {{GULP}: {C}apabilities and prospects},}\ }\href {\doibase doi:10.1524/zkri.220.5.552.65070} {\bibfield  {journal} {\bibinfo  {journal} {Z. Kristallogr. Cryst. Mater.}\ }\textbf {\bibinfo {volume} {220}},\ \bibinfo {pages} {552--554} (\bibinfo {year} {2005})}\BibitemShut {NoStop}%
\bibitem [{\citenamefont {Kresse}\ and\ \citenamefont {Hafner}(1993)}]{Kresse:1993ty}%
  \BibitemOpen
  \bibfield  {author} {\bibinfo {author} {\bibfnamefont {G.}~\bibnamefont {Kresse}}\ and\ \bibinfo {author} {\bibfnamefont {J.}~\bibnamefont {Hafner}},\ }\bibfield  {title} {\enquote {\bibinfo {title} {{A}b initio molecular dynamics for liquid metals},}\ }\href {\doibase 10.1103/physrevb.47.558} {\bibfield  {journal} {\bibinfo  {journal} {Phys. Rev. B}\ }\textbf {\bibinfo {volume} {47}},\ \bibinfo {pages} {558--561} (\bibinfo {year} {1993})}\BibitemShut {NoStop}%
\bibitem [{\citenamefont {Kresse}\ and\ \citenamefont {Hafner}(1994)}]{Kresse:1994us}%
  \BibitemOpen
  \bibfield  {author} {\bibinfo {author} {\bibfnamefont {G.}~\bibnamefont {Kresse}}\ and\ \bibinfo {author} {\bibfnamefont {J.}~\bibnamefont {Hafner}},\ }\bibfield  {title} {\enquote {\bibinfo {title} {{A}b initio molecular-dynamics simulation of the liquid-metal - amorphous-semiconductor transition in germanium},}\ }\href {\doibase 10.1103/physrevb.49.14251} {\bibfield  {journal} {\bibinfo  {journal} {Phys. Rev. B}\ }\textbf {\bibinfo {volume} {49}},\ \bibinfo {pages} {14251--14269} (\bibinfo {year} {1994})}\BibitemShut {NoStop}%
\bibitem [{\citenamefont {Kresse}\ and\ \citenamefont {Furthm{\"u}ller}(1996)}]{Kresse:1996vk}%
  \BibitemOpen
  \bibfield  {author} {\bibinfo {author} {\bibfnamefont {G.}~\bibnamefont {Kresse}}\ and\ \bibinfo {author} {\bibfnamefont {J.}~\bibnamefont {Furthm{\"u}ller}},\ }\bibfield  {title} {\enquote {\bibinfo {title} {{E}fficiency of ab-initio total energy calculations for metals and semiconductors using a plane-wave basis set},}\ }\href {\doibase 10.1016/0927-0256(96)00008-0} {\bibfield  {journal} {\bibinfo  {journal} {Comput. Mater. Sci.}\ }\textbf {\bibinfo {volume} {6}},\ \bibinfo {pages} {15--50} (\bibinfo {year} {1996})}\BibitemShut {NoStop}%
\bibitem [{\citenamefont {Kresse}(1996)}]{Kresse:1996vf}%
  \BibitemOpen
  \bibfield  {author} {\bibinfo {author} {\bibfnamefont {G.}~\bibnamefont {Kresse}},\ }\bibfield  {title} {\enquote {\bibinfo {title} {{E}fficient iterative schemes for ab initio total-energy calculations using a plane-wave basis set},}\ }\href {\doibase 10.1103/physrevb.54.11169} {\bibfield  {journal} {\bibinfo  {journal} {Phys. Rev. B}\ }\textbf {\bibinfo {volume} {54}},\ \bibinfo {pages} {11169--11186} (\bibinfo {year} {1996})}\BibitemShut {NoStop}%
\bibitem [{\citenamefont {Perdew}\ \emph {et~al.}(1996)\citenamefont {Perdew}, \citenamefont {Burke},\ and\ \citenamefont {Ernzerhof}}]{PBE}%
  \BibitemOpen
  \bibfield  {author} {\bibinfo {author} {\bibfnamefont {J.~P.}\ \bibnamefont {Perdew}}, \bibinfo {author} {\bibfnamefont {K.}~\bibnamefont {Burke}}, \ and\ \bibinfo {author} {\bibfnamefont {M.}~\bibnamefont {Ernzerhof}},\ }\bibfield  {title} {\enquote {\bibinfo {title} {{G}eneralized gradient approximation made simple},}\ }\href {\doibase 10.1103/PhysRevLett.77.3865} {\bibfield  {journal} {\bibinfo  {journal} {Phys. Rev. Lett.}\ }\textbf {\bibinfo {volume} {77}},\ \bibinfo {pages} {3865--3868} (\bibinfo {year} {1996})}\BibitemShut {NoStop}%
\bibitem [{\citenamefont {Bl{\"o}chl}(1994)}]{Blochl:1994uk}%
  \BibitemOpen
  \bibfield  {author} {\bibinfo {author} {\bibfnamefont {P.}~\bibnamefont {Bl{\"o}chl}},\ }\bibfield  {title} {\enquote {\bibinfo {title} {{P}rojector augmented-wave method},}\ }\href {\doibase 10.1103/PhysRevB.50.} {\bibfield  {journal} {\bibinfo  {journal} {Phys. Rev. B}\ }\textbf {\bibinfo {volume} {50}},\ \bibinfo {pages} {17953} (\bibinfo {year} {1994})}\BibitemShut {NoStop}%
\bibitem [{\citenamefont {Kresse}\ and\ \citenamefont {Joubert}(1999)}]{Kresse:1999wc}%
  \BibitemOpen
  \bibfield  {author} {\bibinfo {author} {\bibfnamefont {G.}~\bibnamefont {Kresse}}\ and\ \bibinfo {author} {\bibfnamefont {D.}~\bibnamefont {Joubert}},\ }\bibfield  {title} {\enquote {\bibinfo {title} {{F}rom ultrasoft pseudopotentials to the projector augmented-wave method},}\ }\href {\doibase 10.1103/physrevb.59.1758} {\bibfield  {journal} {\bibinfo  {journal} {Phys. Rev. B}\ }\textbf {\bibinfo {volume} {59}},\ \bibinfo {pages} {1758 1775} (\bibinfo {year} {1999})}\BibitemShut {NoStop}%
\bibitem [{\citenamefont {Monkhorst}\ and\ \citenamefont {Pack}(1976)}]{monkhorst1976special}%
  \BibitemOpen
  \bibfield  {author} {\bibinfo {author} {\bibfnamefont {H.~J.}\ \bibnamefont {Monkhorst}}\ and\ \bibinfo {author} {\bibfnamefont {J.~D.}\ \bibnamefont {Pack}},\ }\bibfield  {title} {\enquote {\bibinfo {title} {{S}pecial points for {B}rillouin-zone integrations},}\ }\href {\doibase 10.1103/PhysRevB.13.5188} {\bibfield  {journal} {\bibinfo  {journal} {Phys. Rev. B}\ }\textbf {\bibinfo {volume} {13}},\ \bibinfo {pages} {5188} (\bibinfo {year} {1976})}\BibitemShut {NoStop}%
\bibitem [{\citenamefont {K{\"u}hne}\ \emph {et~al.}(2020)\citenamefont {K{\"u}hne}, \citenamefont {Iannuzzi}, \citenamefont {Del~Ben}, \citenamefont {Rybkin}, \citenamefont {Seewald}, \citenamefont {Stein}, \citenamefont {Laino}, \citenamefont {Khaliullin}, \citenamefont {Sch{\"u}tt}, \citenamefont {Schiffmann}, \citenamefont {Golze}, \citenamefont {Wilhelm}, \citenamefont {Chulkov}, \citenamefont {Bani-Hashemian}, \citenamefont {Weber}, \citenamefont {Bor{\v s}tnik}, \citenamefont {Taillefumier}, \citenamefont {Jakobovits}, \citenamefont {Lazzaro}, \citenamefont {Pabst}, \citenamefont {M{\"u}ller}, \citenamefont {Schade}, \citenamefont {Guidon}, \citenamefont {Andermatt}, \citenamefont {Holmberg}, \citenamefont {Schenter}, \citenamefont {Hehn}, \citenamefont {Bussy}, \citenamefont {Belleflamme}, \citenamefont {Tabacchi}, \citenamefont {Gl{\"o}{\ss}}, \citenamefont {Lass}, \citenamefont {Bethune}, \citenamefont {Mundy}, \citenamefont {Plessl}, \citenamefont {Watkins}, \citenamefont {VandeVondele}, \citenamefont {Krack},\ and\ \citenamefont {Hutter}}]{Kuehne2020}%
  \BibitemOpen
  \bibfield  {author} {\bibinfo {author} {\bibfnamefont {Th.~D.}\ \bibnamefont {K{\"u}hne}}, \bibinfo {author} {\bibfnamefont {M.}~\bibnamefont {Iannuzzi}}, \bibinfo {author} {\bibfnamefont {M.}~\bibnamefont {Del~Ben}}, \bibinfo {author} {\bibfnamefont {V.~V.}\ \bibnamefont {Rybkin}}, \bibinfo {author} {\bibfnamefont {P.}~\bibnamefont {Seewald}}, \bibinfo {author} {\bibfnamefont {F.}~\bibnamefont {Stein}}, \bibinfo {author} {\bibfnamefont {T.}~\bibnamefont {Laino}}, \bibinfo {author} {\bibfnamefont {R.~Z.}\ \bibnamefont {Khaliullin}}, \bibinfo {author} {\bibfnamefont {O.}~\bibnamefont {Sch{\"u}tt}}, \bibinfo {author} {\bibfnamefont {F.}~\bibnamefont {Schiffmann}}, \bibinfo {author} {\bibfnamefont {D.}~\bibnamefont {Golze}}, \bibinfo {author} {\bibfnamefont {J.}~\bibnamefont {Wilhelm}}, \bibinfo {author} {\bibfnamefont {S.}~\bibnamefont {Chulkov}}, \bibinfo {author} {\bibfnamefont {M.~H.}\ \bibnamefont {Bani-Hashemian}}, \bibinfo {author} {\bibfnamefont {V.}~\bibnamefont {Weber}}, \bibinfo {author} {\bibfnamefont {U.}~\bibnamefont {Bor{\v s}tnik}}, \bibinfo {author} {\bibfnamefont {M.}~\bibnamefont {Taillefumier}}, \bibinfo {author} {\bibfnamefont {A.~S.}\ \bibnamefont {Jakobovits}}, \bibinfo {author} {\bibfnamefont {A.}~\bibnamefont {Lazzaro}}, \bibinfo {author} {\bibfnamefont {H.}~\bibnamefont {Pabst}}, \bibinfo {author} {\bibfnamefont {T.}~\bibnamefont {M{\"u}ller}}, \bibinfo {author} {\bibfnamefont {R.}~\bibnamefont {Schade}}, \bibinfo {author} {\bibfnamefont {M.}~\bibnamefont {Guidon}}, \bibinfo {author} {\bibfnamefont {S.}~\bibnamefont {Andermatt}}, \bibinfo {author} {\bibfnamefont {N.}~\bibnamefont {Holmberg}}, \bibinfo {author} {\bibfnamefont {G.~K.}\ \bibnamefont {Schenter}}, \bibinfo {author} {\bibfnamefont {A.}~\bibnamefont {Hehn}}, \bibinfo {author} {\bibfnamefont {A.}~\bibnamefont {Bussy}}, \bibinfo {author} {\bibfnamefont {F.}~\bibnamefont {Belleflamme}}, \bibinfo {author} {\bibfnamefont {G.}~\bibnamefont {Tabacchi}}, \bibinfo {author} {\bibfnamefont {A.}~\bibnamefont {Gl{\"o}{\ss}}}, \bibinfo
  {author} {\bibfnamefont {M.}~\bibnamefont {Lass}}, \bibinfo {author} {\bibfnamefont {I.}~\bibnamefont {Bethune}}, \bibinfo {author} {\bibfnamefont {Ch.~J.}\ \bibnamefont {Mundy}}, \bibinfo {author} {\bibfnamefont {Ch.}\ \bibnamefont {Plessl}}, \bibinfo {author} {\bibfnamefont {M.}~\bibnamefont {Watkins}}, \bibinfo {author} {\bibfnamefont {J.}~\bibnamefont {VandeVondele}}, \bibinfo {author} {\bibfnamefont {M.}~\bibnamefont {Krack}}, \ and\ \bibinfo {author} {\bibfnamefont {J.}~\bibnamefont {Hutter}},\ }\bibfield  {title} {\enquote {\bibinfo {title} {{CP2K}: {A}n electronic structure and molecular dynamics software package - {Q}uickstep: {E}fficient and accurate electronic structure calculations},}\ }\href {\doibase 10.1063/5.0007045} {\bibfield  {journal} {\bibinfo  {journal} {J. Chem. Phys.}\ }\textbf {\bibinfo {volume} {152}},\ \bibinfo {pages} {194103} (\bibinfo {year} {2020})}\BibitemShut {NoStop}%
\bibitem [{\citenamefont {Goedecker}\ \emph {et~al.}(1996)\citenamefont {Goedecker}, \citenamefont {Teter},\ and\ \citenamefont {Hutter}}]{Goedecker1996}%
  \BibitemOpen
  \bibfield  {author} {\bibinfo {author} {\bibfnamefont {S.}~\bibnamefont {Goedecker}}, \bibinfo {author} {\bibfnamefont {M.}~\bibnamefont {Teter}}, \ and\ \bibinfo {author} {\bibfnamefont {J.}~\bibnamefont {Hutter}},\ }\bibfield  {title} {\enquote {\bibinfo {title} {{S}eparable dual-space {G}aussian pseudopotentials},}\ }\href {\doibase 10.1103/PhysRevB.54.1703} {\bibfield  {journal} {\bibinfo  {journal} {Phys. Rev. B}\ }\textbf {\bibinfo {volume} {54}},\ \bibinfo {pages} {1703--1710} (\bibinfo {year} {1996})}\BibitemShut {NoStop}%
\bibitem [{\citenamefont {VandeVondele}\ and\ \citenamefont {Hutter}(2007)}]{VandeVondele2007}%
  \BibitemOpen
  \bibfield  {author} {\bibinfo {author} {\bibfnamefont {J.}~\bibnamefont {VandeVondele}}\ and\ \bibinfo {author} {\bibfnamefont {J.}~\bibnamefont {Hutter}},\ }\bibfield  {title} {\enquote {\bibinfo {title} {{G}aussian basis sets for accurate calculations on molecular systems in gas and condensed phases},}\ }\href {\doibase 10.1063/1.2770708} {\bibfield  {journal} {\bibinfo  {journal} {J. Chem. Phys.}\ }\textbf {\bibinfo {volume} {127}},\ \bibinfo {pages} {114105} (\bibinfo {year} {2007})}\BibitemShut {NoStop}%
\bibitem [{\citenamefont {Freysoldt}\ \emph {et~al.}(2014)\citenamefont {Freysoldt}, \citenamefont {Grabowski}, \citenamefont {Hickel}, \citenamefont {Neugebauer}, \citenamefont {Kresse}, \citenamefont {Janotti},\ and\ \citenamefont {Van~de Walle}}]{freysoldt2014first}%
  \BibitemOpen
  \bibfield  {author} {\bibinfo {author} {\bibfnamefont {C.}~\bibnamefont {Freysoldt}}, \bibinfo {author} {\bibfnamefont {B.}~\bibnamefont {Grabowski}}, \bibinfo {author} {\bibfnamefont {T.}~\bibnamefont {Hickel}}, \bibinfo {author} {\bibfnamefont {J.}~\bibnamefont {Neugebauer}}, \bibinfo {author} {\bibfnamefont {G.}~\bibnamefont {Kresse}}, \bibinfo {author} {\bibfnamefont {A.}~\bibnamefont {Janotti}}, \ and\ \bibinfo {author} {\bibfnamefont {C.~G.}\ \bibnamefont {Van~de Walle}},\ }\bibfield  {title} {\enquote {\bibinfo {title} {{F}irst-principles calculations for point defects in solids},}\ }\href {\doibase 10.1103/RevModPhys.86.253} {\bibfield  {journal} {\bibinfo  {journal} {Rev. Mod. Phys.}\ }\textbf {\bibinfo {volume} {86}},\ \bibinfo {pages} {253} (\bibinfo {year} {2014})}\BibitemShut {NoStop}%
\bibitem [{\citenamefont {Ruiz~Caridad}\ \emph {et~al.}(2022)\citenamefont {Ruiz~Caridad}, \citenamefont {Erni}, \citenamefont {Vogel},\ and\ \citenamefont {Rossell}}]{Ruiz22}%
  \BibitemOpen
  \bibfield  {author} {\bibinfo {author} {\bibfnamefont {A.}~\bibnamefont {Ruiz~Caridad}}, \bibinfo {author} {\bibfnamefont {R.}~\bibnamefont {Erni}}, \bibinfo {author} {\bibfnamefont {A.}~\bibnamefont {Vogel}}, \ and\ \bibinfo {author} {\bibfnamefont {M.~D.}\ \bibnamefont {Rossell}},\ }\bibfield  {title} {\enquote {\bibinfo {title} {{A}pplications of a novel electron energy filter combined with a hybrid-pixel direct electron detector for the analysis of functional oxides by stem-eels and energy-filtered imaging},}\ }\href {\doibase 10.1016/j.micron.2022.103331} {\bibfield  {journal} {\bibinfo  {journal} {Micron}\ }\textbf {\bibinfo {volume} {160}},\ \bibinfo {pages} {103331} (\bibinfo {year} {2022})}\BibitemShut {NoStop}%
\bibitem [{\citenamefont {Campanini}\ \emph {et~al.}(2018)\citenamefont {Campanini}, \citenamefont {Erni}, \citenamefont {Yang}, \citenamefont {Ramesh},\ and\ \citenamefont {Rossell}}]{Campanini2018}%
  \BibitemOpen
  \bibfield  {author} {\bibinfo {author} {\bibfnamefont {M.}~\bibnamefont {Campanini}}, \bibinfo {author} {\bibfnamefont {R.}~\bibnamefont {Erni}}, \bibinfo {author} {\bibfnamefont {C.-H.}\ \bibnamefont {Yang}}, \bibinfo {author} {\bibfnamefont {R.}~\bibnamefont {Ramesh}}, \ and\ \bibinfo {author} {\bibfnamefont {M.~D.}\ \bibnamefont {Rossell}},\ }\bibfield  {title} {\enquote {\bibinfo {title} {{P}eriodic giant polarization gradients in doped \ce{BiFeO3} thin films},}\ }\href {\doibase 10.1021/acs.nanolett.7b03817} {\bibfield  {journal} {\bibinfo  {journal} {Nano Lett.}\ }\textbf {\bibinfo {volume} {18}},\ \bibinfo {pages} {171--724} (\bibinfo {year} {2018})}\BibitemShut {NoStop}%
\bibitem [{\citenamefont {Muller}\ \emph {et~al.}(2004)\citenamefont {Muller}, \citenamefont {Nakagawa}, \citenamefont {Ohtomo}, \citenamefont {Grazul},\ and\ \citenamefont {Hwang}}]{Muller2004}%
  \BibitemOpen
  \bibfield  {author} {\bibinfo {author} {\bibfnamefont {D.~A.}\ \bibnamefont {Muller}}, \bibinfo {author} {\bibfnamefont {N.}~\bibnamefont {Nakagawa}}, \bibinfo {author} {\bibfnamefont {A.}~\bibnamefont {Ohtomo}}, \bibinfo {author} {\bibfnamefont {J.~L.}\ \bibnamefont {Grazul}}, \ and\ \bibinfo {author} {\bibfnamefont {H.~Y.}\ \bibnamefont {Hwang}},\ }\bibfield  {title} {\enquote {\bibinfo {title} {{A}tomic-scale imaging of nanoengineered oxygen vacancy profiles in \ce{SrTiO3}},}\ }\href {\doibase 10.1038/nature02756} {\bibfield  {journal} {\bibinfo  {journal} {Nature}\ }\textbf {\bibinfo {volume} {430}},\ \bibinfo {pages} {657--661} (\bibinfo {year} {2004})}\BibitemShut {NoStop}%
\bibitem [{\citenamefont {Aschauer}\ \emph {et~al.}(2015)\citenamefont {Aschauer}, \citenamefont {Vonr{\"u}ti},\ and\ \citenamefont {Spaldin}}]{Aschauer2015}%
  \BibitemOpen
  \bibfield  {author} {\bibinfo {author} {\bibfnamefont {U.}~\bibnamefont {Aschauer}}, \bibinfo {author} {\bibfnamefont {N.}~\bibnamefont {Vonr{\"u}ti}}, \ and\ \bibinfo {author} {\bibfnamefont {N.~A.}\ \bibnamefont {Spaldin}},\ }\bibfield  {title} {\enquote {\bibinfo {title} {{E}ffect of epitaxial strain on cation and anion vacancy formation in \ce{MnO}},}\ }\href {\doibase 10.1103/PhysRevB.92.054103} {\bibfield  {journal} {\bibinfo  {journal} {Phys. Rev. B}\ }\textbf {\bibinfo {volume} {92}},\ \bibinfo {pages} {054103} (\bibinfo {year} {2015})}\BibitemShut {NoStop}%
\bibitem [{\citenamefont {Abbate}\ \emph {et~al.}(1991)\citenamefont {Abbate}, \citenamefont {de~Groot}, \citenamefont {Fuggle}, \citenamefont {Fujimori}, \citenamefont {Tokura}, \citenamefont {Fujishima}, \citenamefont {Strebel}, \citenamefont {Domke}, \citenamefont {Kaindl}, \citenamefont {van Elp}, \citenamefont {Thole}, \citenamefont {Sawatzky}, \citenamefont {Sacchi},\ and\ \citenamefont {Tsuda}}]{Abbate1991}%
  \BibitemOpen
  \bibfield  {author} {\bibinfo {author} {\bibfnamefont {M.}~\bibnamefont {Abbate}}, \bibinfo {author} {\bibfnamefont {F.~M.~F.}\ \bibnamefont {de~Groot}}, \bibinfo {author} {\bibfnamefont {J.~C.}\ \bibnamefont {Fuggle}}, \bibinfo {author} {\bibfnamefont {A.}~\bibnamefont {Fujimori}}, \bibinfo {author} {\bibfnamefont {Y.}~\bibnamefont {Tokura}}, \bibinfo {author} {\bibfnamefont {Y.}~\bibnamefont {Fujishima}}, \bibinfo {author} {\bibfnamefont {O.}~\bibnamefont {Strebel}}, \bibinfo {author} {\bibfnamefont {M.}~\bibnamefont {Domke}}, \bibinfo {author} {\bibfnamefont {G.}~\bibnamefont {Kaindl}}, \bibinfo {author} {\bibfnamefont {J.}~\bibnamefont {van Elp}}, \bibinfo {author} {\bibfnamefont {B.~T.}\ \bibnamefont {Thole}}, \bibinfo {author} {\bibfnamefont {G.~A.}\ \bibnamefont {Sawatzky}}, \bibinfo {author} {\bibfnamefont {M.}~\bibnamefont {Sacchi}}, \ and\ \bibinfo {author} {\bibfnamefont {N.}~\bibnamefont {Tsuda}},\ }\bibfield  {title} {\enquote {\bibinfo {title} {{S}oft-x-ray-absorption studies of the location of extra charges induced by substitution in controlled-valence materials},}\ }\href {\doibase 10.1103/PhysRevB.44.5419} {\bibfield  {journal} {\bibinfo  {journal} {Phys. Rev. B}\ }\textbf {\bibinfo {volume} {44}},\ \bibinfo {pages} {5419--5422} (\bibinfo {year} {1991})}\BibitemShut {NoStop}%
\bibitem [{\citenamefont {Ricca}\ \emph {et~al.}(2020{\natexlab{b}})\citenamefont {Ricca}, \citenamefont {Timrov}, \citenamefont {Cococcioni}, \citenamefont {Marzari},\ and\ \citenamefont {Aschauer}}]{Ricca2020_sto}%
  \BibitemOpen
  \bibfield  {author} {\bibinfo {author} {\bibfnamefont {C.}~\bibnamefont {Ricca}}, \bibinfo {author} {\bibfnamefont {I.}~\bibnamefont {Timrov}}, \bibinfo {author} {\bibfnamefont {M.}~\bibnamefont {Cococcioni}}, \bibinfo {author} {\bibfnamefont {N.}~\bibnamefont {Marzari}}, \ and\ \bibinfo {author} {\bibfnamefont {U.}~\bibnamefont {Aschauer}},\ }\bibfield  {title} {\enquote {\bibinfo {title} {{S}elf-consistent {DFT+U+V} study of oxygen vacancies in \ce{SrTiO3}},}\ }\href {\doibase 10.1103/PhysRevResearch.2.023313} {\bibfield  {journal} {\bibinfo  {journal} {Phys. Rev. Research}\ }\textbf {\bibinfo {volume} {2}},\ \bibinfo {pages} {023313} (\bibinfo {year} {2020}{\natexlab{b}})}\BibitemShut {NoStop}%
\end{thebibliography}%

\end{document}